Reduced model for deep bed filtration of binary highly heterogeneous colloids

Nastaran Khazali[a], Thomas Russell[a], Gabriel Malgaresi[a], Chengyuan Xu[b], Pavel Bedrikovetsky[a]

[a] School of Chemical Engineering, University of Adelaide, SA, Australia

[b] State Key Laboratory of Oil and Gas Reservoir Geology and Exploitation, Southwest Petroleum University, Chengdu 610500, China

## Abstract

Hyper-exponential retention profiles (HERPs) are often observed during laboratory tests on colloidal and nano-suspension transport in porous media. The aim of this work is an extension of the traditional model for suspended particle transport (deep bed filtration) to match HERPs. Interpreting particle capture in the framework of active mass law for the "chemical reaction" particle-rock, we substitute suspension concentration in the traditional expression for capture rate by non-linear "activity" function, called the suspension function. The non-linear suspension function is derived from Margules' activity model for excess Gibbs free energy. This suspension function is shown to match that derived from multicomponent colloidal transport with good agreement, despite having one less unknown parameter. Empirical formulae are presented to convert between the two unknown parameters of the multicomponent model and the Margules exponent. Comparison between the new model and laboratory coreflooding data shows good agreement, even in cases with distinctly hyper-exponential retention profiles, where the classical model fails to reproduce the laboratory data.

## 1. Introduction

Colloidal and nano-suspension flow in porous media is of fundamental importance to environmental, chemical, civil and petroleum engineering. Deep bed filtration (DBF) includes propagation of viruses, bacteria and other contaminants in ground waters, disposal of industrial wastes in aquifers, industrial filtering of technical fluids, transport of nanoparticles, injection of water with solid and liquid particles in oilfields during waterflooding, size-exclusion chromatography, and technical fluid invasion during well drilling and completion [1-11]. The major features of the processes are particle transport and capture by the porous medium, and the consequential permeability decline.

The mathematical model for colloidal and nano- suspension transport in porous media includes mass balance for the suspended and retained particles, expression for particle capture rate, and Darcy´s law that accounts for permeability decline during particle retention [12, 13]

$$\phi\frac{\partial c}{\partial t}+U\frac{\partial c}{\partial x}=-\frac{\partial s}{\partial t} \tag{1}$$

$$\frac{\partial s}{\partial t}=h(s)\lambda cU \tag{2}$$

$$U=-\frac{k_0}{\mu(1+\beta s)}\frac{\partial p}{\partial x} \tag{3}$$

Here $\phi$ is the porosity, $U$ is the Darcy's flow velocity, $c$ is the suspension concentration, $s$ is the retained concentration, $h(s)$ is the filtration function, measuring the change in capture rate with changing concentrations of vacancies within the porous media, $\lambda$ is the filtration coefficient, $p$ is the pressure, $k_0$ is the initial permeability, $\beta$ is the formation damage coefficient, and $\mu$ is the carrier water viscosity. The suspension concentration is defined as the number of particles per unit volume of the fluid, and the retained concentration is equal to the number of particles per unit of the rock volume. The filtration coefficient $\lambda$ is the probability of particle capture per unit length of its trajectory [14-16]. The relationship between the damaged and undamaged permeability follows from a first order Taylor's series approximation.

Fig. 1 presents multiple particle capture mechanisms that may occur simultaneously during flow of colloidal and nano-suspensions in porous media: including straining, attachment via electrostatic forces, and bridging [17-19]. Particle capture is analogous to a chemical reaction between the particle and vacancy. Vacancy for size exclusion is a pore throats that is smaller than the particle [13]. Vacancy for attachment is a site on the rock surface that attracts the particle by electrostatic forces. Kinetics equation (2) corresponds to the total particle capture rate by the rock.

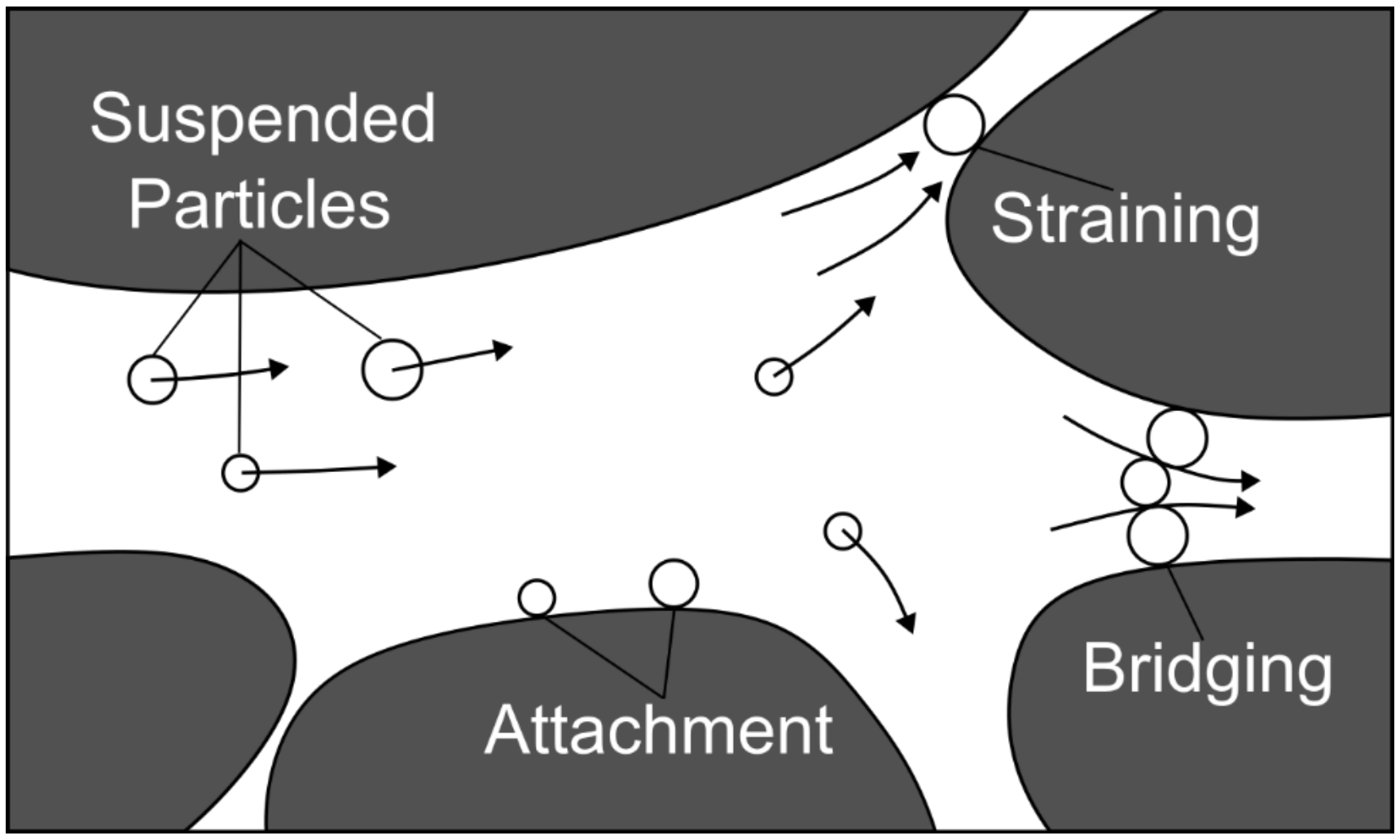

Fig. 1. Schematic for multi-mechanism particle capture in porous media

The initial and boundary conditions that correspond to colloid injection with constant concentration into a clean bed are:

$$\begin{aligned} t=0:\ & c=s=0 \\ x=0:\ & c=c^0 \end{aligned} \tag{4}$$

The most common form of the filtration function, $h(s)$, is the Langmuir blocking function [13]:

$$h(s)=\left(1-\frac{s}{s_m}\right) \tag{5}$$

The blocking function corresponds to mono-layer particle attachment to the rock surface. It also describes suspension-colloidal transport with mono-sized particles through porous media with two-size pores, where the particles move via large pores and strain the small pores [16]; here $s_m$ is the initial concentration of small pores. However, in the case where pore plugging increases the particle path tortuosity and lengthens the trajectory, the particle straining increases the capture probability; matching of the laboratory data yields a negative value of the phenomenological constant $s_m$.

In the clean bed case, the initial vacancy concentration is $s_m$, and concentration of remaining vacancies is $s_m$-$s$. So, kinetics equation (5) is analogous to active mass law in chemical reaction kinetics with both stoichiometric coefficients equal to one, as if one vacancy "reacts" with the captured particle [20]. For small concentrations of suspended particles and vacancies, the particle capture is a result of simultaneous appearance of a suspended particle and a vacancy in the same reference volume. In this case, the simultaneous appearances are independent events, so the probability of "reaction" is equal to product of the individual probabilities. Therefore, the capture rate is proportional to the product of reacting species concentrations $c(s_m\text{-}s)$. This form of the active mass law is valid for low concentrations of both reagents [21].

Experiments studying colloidal transport and capture typically measure the outlet suspended concentration, $c(1,t)$, and the retained concentration profile, $s(x,t^*)$. The solution of system ((1),(2)) exhibits retention profiles that decrease exponentially (or less sharply when $h(s)$ is a decreasing function). However, numerous corefloods by suspensions and colloids exhibit hyper-exponential retention profiles (HERPs) [8, 22-30]. Those profiles correspond to retention rate decrease along the core. The traditional filtration model does not match the HERPs [31]. In particular, Tufenkji and Elimelech [23], [32] highlight that the measured retained concentrations are in marked discrepancy with those predicted by Eq. (2). To match the retained profiles, Tufenkji and Elimelech [8] assumed probabilistic distribution of the filtration coefficient. This model assumes bimodal distribution with five constants, corresponding to fast and slow capture due to favourable and unfavourable particle-surface interactions. The proposed stochastic model successfully matches the retention profiles using those five constants.

Schijven, Hassanizadeh and de Bruin [6] proposed a two-capture model for colloidal-suspension transport in porous media. The model uses four constants for tuning. This model successfully matches either HERPs or BTC; simultaneous match of two curves cannot be achieved [31].

Another way around the matching of HERPs was developed by Bradford, Simunek, Bettahar, van Genuchten and Yates [33], where the filtration function for straining is depth-dependent, i.e. $h=h(s,x)$. Power-law is assumed for this depth-dependency. The model also included kinetics of simultaneous attachment and detachment. The model successfully matched BTCs and HERPs simultaneously. Currently, this model is widely used for matching of numerous laboratory data sets [17, 30, 34, 35]. However, Goldberg, Scheringer, Bucheli and Hungerbühler [31] suggested that $x$-dependency of the filtration function must be justified with more physical background.

Numerous authors attribute HERPs to distributed properties of colloidal particles [22, 23, 25, 32]. In particular, Tong and Johnson [22] explain HERPs by multiple populations in the colloidal fluids. The same size particles can have different surface charges and zeta-potentials, so the particles with high filtration coefficients retain near to the inlet, while those with lower filtration coefficient move further in the core. The total of two exponential RPs with highly different decrements is non-exponential.

Yuan and Shapiro [36] also discuss multi-component colloids, extending the random walk model with stochastically distributed lengths and durations of jumps [37], in order to match simultaneously retention profile and BTC curve. Averaging of the Master equation with distributed jump lengths and flight times results in elliptic transport equation. The filtration coefficient is assumed to be dependent of a particle property, like size, shape, surface charge, etc. Introduction of probabilistic distributions of the filtration coefficient and of the particle flight time into Master equation yields system of elliptic transport equations, where each equation corresponds to a particle population. The derived "integral" system successfully matches HERPs and BTCs simultaneously.

Exact averaging of a stochastic system results in representation of a distributed parameter by a number (function). For example, jump length distribution in random walks and consequent averaging result in diffusion coefficient [38]. Averaging of stochastically distributed lengths and durations of jumps leads to diffusion and timely-diffusion coefficients [37]. Discretisation of a wide filtration-coefficient distribution in the system with distributed lengths and durations of jumps requires numerous values for filtration coefficient, and leads to solution of the system containing the same number of elliptic equations. This results in cumbersome and computer time consuming calculations for direct runs and for iterative data-matching procedures. Averaging of this system would significantly simplify solutions of forward and inverse problems.

In the current work we develop a new formulation for deep bed filtration in order to match HERPs. The new formulation further develops the analogy between chemical reaction kinetics and colloidal filtration in porous media by leveraging Margules' activity model to derive a non-linear suspension function. This new suspension function is compared with that derived from multicomponent colloidal flow and compared with experimental data to assess its ability to capture the underlying characteristics which give rise to HERPs. The new model has the advantage that it contains one less unknown parameter compared to even the simplest multicomponent colloid model, thus decreasing the information requirements for characterisation of filtration systems. By demonstrating the approximate equivalence between the multicomponent and Margules models, we improve on the practical impact of the model whilst maintaining the strong physical basis of the multicomponent model.

The structure of the text is as follows. Section 2 presents the formulation of the Margules and multicomponent colloid models. Section 3 derives the analytical solution for colloidal transport for any arbitrary suspension function for both Langmuir and constant filtration functions. In Section 4 we compare the Margules and binary particle models and derive an approximate equation to convert between their unknown coefficients. Section 5 presents matching of laboratory data with the Margules, binary, and classical models. Section 6 presents discussions and Section 7 concludes the work.

## 2. Formulation of the model for non-linear filtration of colloids

In this section we will present the formulation of the model for suspension colloidal flow through porous media with a non-linear suspension function. Following the presentation of the general system for colloidal flow, we then discuss how a non-linear suspension function can arise from multi population particle filtration, or by using an analogy to activity coefficients during chemical reactions.

For all equations further in the text, we use the following set of dimensionless variables:

$$x=\frac{x}{L},t=\frac{1}{\phi L}\int_0^t U(y)\,dy,c\rightarrow\frac{c}{c^0},s\rightarrow\frac{s}{\phi c^0},p\rightarrow\frac{k_0 p}{UL\mu},s_m\rightarrow\frac{s_m}{\phi c^0},f(c)\rightarrow\frac{Lf(c)}{c^0} \tag{6}$$

where $L$ is the system length, and $c^0$ is the initial injected concentration.

## 2.1. Governing equations for non-linear filtration of colloids

The governing equations for colloidal flow in porous media include the mass balance on the particles, as well as the equation for the capture rate:

$$\frac{\partial c}{\partial t}+\frac{\partial c}{\partial x}=-\frac{\partial s}{\partial t} \tag{7}$$

$$\frac{\partial s}{\partial t}=h(s)f(c) \tag{8}$$

$$1=-\frac{1}{(1+\beta s)}\frac{\partial p}{\partial x} \tag{9}$$

where $h(s)$ is the filtration function, and $f(c)$ is the suspension function. The filtration function describes the evolution of the capture rate as more particles occupy capture sites within the porous media. Typically this function decreases monotonically, corresponding to a 'depletion' of capture sites over time [39], however in some cases $h(s)$ can increase [40], for example when particles find attached particles as more attractive attachment sites than the natural internal surface of the porous media.

The dimensionless initial and boundary conditions are:

$$\begin{aligned} t=0 &: c=s=0 \\ x=0 &: c=1 \end{aligned} \tag{10}$$

Hyper-exponential retention profiles arise from the non-linearity of the suspension function, $f(c)$. Let us now discuss two cases in which such functions can be derived.

## 2.2. Margules activity model

In describing the rate of chemical reactions, we often describe the rate of kinetics as being proportional to the concentration of each reacting species. However, in some cases, the reaction rate varies non-linearly with the reactant concentrations. To this end, we use the concept of 'activity' as an 'effective concentration' such that the reaction rate will be proportional to the activity of each reactant. The activity is defined as:

$a=\gamma x$

where $a$ is the activity, $\gamma_i$ is the activity coefficient, and $x_i$ is the mole fraction of species $i$.

Based on the derivation of the Margules activity coefficients presented in Appendix A, the activities of each component are as follows:

$$a_1 = \gamma_1 x_1 = x_1 \exp(A x_2^2), \quad a_2 = \gamma_2 x_2 = x_2 \exp(A x_1^2) \tag{11}$$

Now let us make an analogy between chemical reaction kinetics and the rate of particle capture. Just as chemical reactions occur due to the coincidence of molecules of two molecules, particle capture occurs due to the physical intersection of a suspended particle and a capture site on the porous surface. To allow the theory to apply to different capture mechanisms, we do not specify whether this capture site corresponds to exposed internal surface area of the porous media, grain-grain contacts, or small pore throats, among others.

Thus, given that a chemical reaction proceeds at a rate determined by the product of the activity of each reactant, we consider the filtration rate to be proportional to the product of the 'activity' associated with the particle and capture site within the porous media. Based on this analogy we may set:

$$x_1 = c, \qquad x_2 = 1 - \frac{s}{s_m} \tag{12}$$

Thus from the two equations in (11) we have:

$$a_1 = c \exp\left( A \left( 1 - \frac{s}{s_m} \right)^2 \right), \qquad a_2 = \left( 1 - \frac{s}{s_m} \right) \exp(Ac^2) \tag{13}$$

Therefore, the expression for the capture rate is:

$$\frac{\partial s}{\partial t} = \lambda c \left( 1 - \frac{s}{s_m} \right) \exp\left( A \left( 1 - \frac{s}{s_m} \right)^2 \right) \exp(Ac^2) \tag{14}$$

where we have included the kinetics coefficient, which for deep bed filtration is the filtration coefficient, $\lambda$.

As we will show in a subsequent section, while an exact solution for colloidal suspension flow exists for any suspension function, *f(c)*, a general formula for a general filtration function, *h(s)* is not available. Solutions are available however for constant filtrations and functions which vary linearly with $s$ (so called Langmuir filtration functions). As such we would like to consider the case in which:

$$\left( 1 - \frac{s}{s_m} \right)^2 \to 0 \tag{15}$$

such that the expression for the capture rate becomes:

$$\frac{\partial s}{\partial t} = \lambda c \left( 1 - \frac{s}{s_m} \right) \exp(Ac^2) \tag{16}$$

In Appendix B, we present a quantitative analysis of the conditions under which using Eq. (16) as an approximation for Eq. (14) are valid. This is performed by comparing the travelling wave solutions for each equation. This analysis shows a wide range under which this approximation is valid, particularly at large times, and thus further in the text we will use the 'simplified' Margules equation (16).

### 2.3. Upscaling of multi population colloidal transport

Bedrikovetsky, Osipov, Kuzmina and Malgaresi [41] have shown that injection of multicomponent particle populations can be averaged resulting in a single mass balance and capture rate equation with a non-linear suspension function which reflects the capture properties of each individual sub population of particles. This model has been further developed to show that it can reproduce both non-monotonic retention profiles [42] as well as HERPs [40]. Consider the injection of multiple particles. For each particle we write the following mass balance and capture rate equation (in dimensionless coordinates):

$$\frac{\partial}{\partial t}\left(c_k + s_k\right) + \frac{\partial}{\partial x}\left(c_k\right) = 0$$

(17)

$$\frac{\partial s_k}{\partial t} = h\left(\sum_{i=1}^{m} s_i\right) F_k\left(c_k\right), \;\; k = 1, 2, ..., N \tag{18}$$

where $F_k(c_k)$ are the suspension functions for each subpopulation $k$.

The initial conditions and boundary conditions are:

$$\begin{aligned} t = 0 &: c_k = s_k = 0 \\ x = 0 &: c_k = c_k^0 \end{aligned} \tag{19}$$

We introduce the total suspended and captured particle concentrations as follows:

$$c = \sum_{k=1}^{N} c_k, \; s = \sum_{k=1}^{N} s_k \tag{20}$$

Based on derivations presented in Appendix C, this averaging leads to Eq. (C9) for the suspension function. Equations (C6) and (C7) determine the $g_k(c)$ functions necessary to evaluate $f(c)$.

Consider small, injected concentrations where the suspension functions are linear, and a binary system with only two injected subpopulations of particles:

$$F_k(c_k) = \lambda_k c_k, \;\; k = 1, 2 \tag{21}$$

Formulae (C3), (C5), (C6) for $N$=2 become:

$$G_k(c_k) = \ln\left[\frac{c_k}{c_k^0}\right]^{\frac{1}{\lambda_k}}, \; c_2 = c_2^0\left[\frac{c_1}{c_1^0}\right]^{\frac{\lambda_2}{\lambda_1}}, \;\; c = c_1 + c_2^0\left[\frac{c_1}{c_1^0}\right]^{\frac{\lambda_2}{\lambda_1}}, \;\; n = \frac{\lambda_1}{\lambda_2}; \; k = 1, 2 \tag{22}$$

Yielding explicit expressions for suspension and occupation functions:

$$f(c)=\lambda_1 g_1(c)+\lambda_2 c_2^0\left[\frac{g_1(c)}{c_1^0}\right]^{\frac{\lambda_2}{\lambda_1}} \tag{23}$$

The binary system contains two additional parameters, the relative injected concentration, $c_1^0$, and $\theta=(\lambda_1-\lambda_2)/\lambda_1$. After normalising the injected concentration, both of these parameters vary between 0 and 1.

## 3. Analytical models for 1D transport of non-linear filtration of colloids

In this section we present the solution of the 1D colloidal suspension flow problem ((7),(8)) subject to initial and boundary conditions (10).

### 3.1. Langmuir filtration function

The most common form of the filtration function is the Langmuir equation (5), commonly found in adsorption applications. For this filtration function, Appendix D presents the derivation of the analytical solution. The solution involves splitting the space ($x$,$t$) into two regions separated by the curve $x=t$. For all values of $x$ ahead of this curve ($x>t$), the presence of the injected particles is not felt. Following the derivation, the concentration of suspended particles on this front is solved by evaluating the integral:

$$\int_{c^-(x)}^{1}\frac{du}{f(u)}=x \tag{24}$$

where $c_1^-(x)$ is the concentration immediately behind the injected particle front.

The suspended concentration for all points behind the front can then be calculated implicitly using:

$$\int_{c^-(x)}^{c}\frac{du}{(1-u)f(u)}=\frac{t-x}{s_m} \tag{25}$$

The strained particle concentration, $s$, can then be evaluated using the expression:

$$s(x,t)=s_m\left[\frac{c(x,t)-c^-(x,x)}{1-c^-(x,x)}\right] \tag{26}$$

For some particular forms of suspension function, we can derive explicit solutions for $c(x,t)$ and $s(x,t)$. For example, for the traditional suspension function, where $f(c)=\lambda c$, if we inject a constant concentration, $c^0(t)=c^0$, the concentration of suspended particles just behind the front can be found by integrating Eq. (24):

$$e^{-x\lambda}=c^-(x) \tag{27}$$

Then the suspended particle concentration can be calculated by integrating Eq. (D16):

$$c=\left[1-\exp\left(-\lambda\frac{\tau}{s_m}\right)+\exp\left(-\lambda\frac{\tau}{s_m}+\lambda x\right)\right]^{-1} \tag{28}$$

Finally the strained concentration can be calculated from Eq. (29) using the solution for $c^{-}(x,x)$ and $c(x,t)$:

$$s(x,t)=s_m\left[1-\exp\left(-\lambda\frac{\tau}{s_m}\right)\right]\left[1-\exp\left(-\lambda\frac{\tau}{s_m}\right)+\exp\left(-\lambda\frac{\tau}{s_m}+\lambda x\right)\right]^{-1} \tag{29}$$

## 3.2. Constant filtration function

When the filtration function is constant ($h(s)=1$) or when $s_m$ is infinity, the concentration of suspended particles behind the front and just behind the front are equal and can be obtained using equation (24). Also, the retained particle concentration can be obtained using equation (26). For the traditional suspension function, where $f(c)=\lambda c$, explicit solutions for $c(x,t)$ and $s(x,t)$ become:

$$c(x)=e^{-\lambda x},\quad s(x,t)=\lambda e^{-\lambda x}t \tag{30}$$

## 3.3. Type curves for Margules suspension function

Using the analytical solution for suspended (25) and retained (26) particle concentrations, we can generate type curves for the Margules suspension function (16) to investigate its behaviour. Fig. 2 presents type curves for the suspended concentration at the outlet (referred to as the breakthrough concentration (BTC)), $c(1,t)$, the retained particle profiles, $s(x,t^*)$ plotted in log coordinates, and the impedance, $J$ (dimensionless pressure drop), $J(t)$. As the value of the Margules exponent, $A$, increases, the intensity of particle capture increases. This increase corresponds to a more non-linear suspension function, which results in Fig. 2b in an increasing hyper-exponentiality of the retention profiles. This also corresponds to a greater curvature in the growth of the impedance.

What we observe in Fig. 2 and is evident from Eq. (16) is that when the Margules exponent, $A$, tends to zero, the filtration rate tends to the classical model (2). Based on the analogy to chemical reactions, this is equivalent to a system reverting to 'ideal' behaviour, where the activity of each component is equal to its mole fraction. This is highlighted in Fig. 2b, where we observe the tendency of the retained concentration towards an exponential profile.

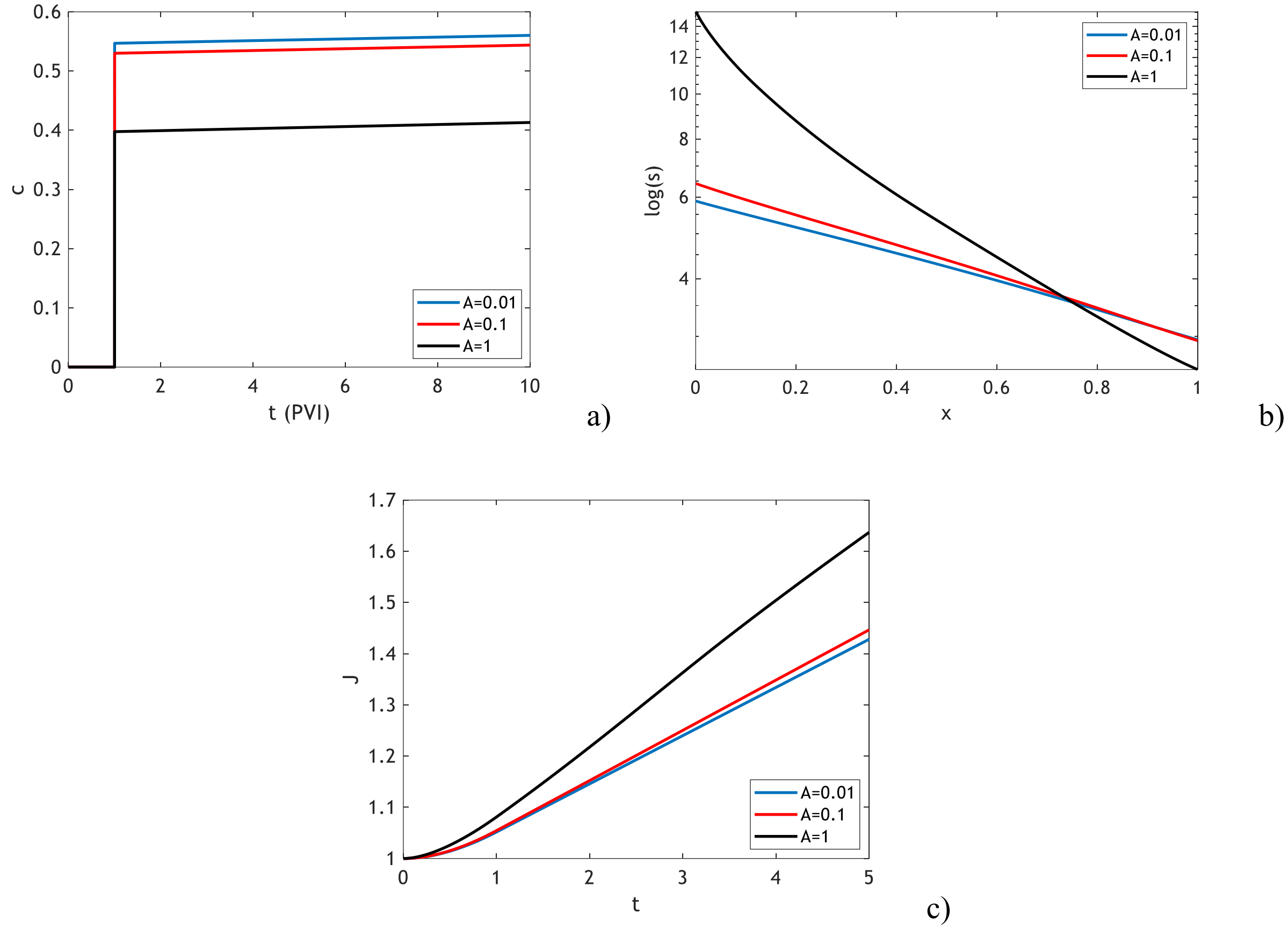

Fig. 2. Type curves and their sensitivity with respect to $A$: a) breakthrough concentrations b) retained concentrations c) pressure drops: $s_m$=100, $\lambda$=0.6, $\beta$=1000, ϕ=0.1, $c_1^0$= $4.15\times10^{-6}$

## 4. Matching the modelling results of multicomponent colloid transport

In this section we show that to a good degree of approximation, the model for colloidal transport using the simplified Margules suspension function matches the behaviour of multicomponent colloid transport. To achieve this, first we attempt to match the suspension functions directly, to assess whether the Margules model can capture the non-linearity in the suspension function of the multicomponent model that ultimately gives rise to non-classical behaviour such as hyper-exponential retention functions. The matching is achieved by performing the following minimisation over the range $c \in [0,1]$:

$$\min_{A}\left(f_B(c)-f_M(c,A,\lambda)\right)^2 \tag{31}$$

Where $f_B(c)$ is the binary suspension function, and $f_M(c,A,\lambda)$ is the Margules suspension function.

The results for three values of $\theta$ are presented in Fig. 3, showing good agreement between the two models. The full set of tuning results are presented in Tables 1-3.

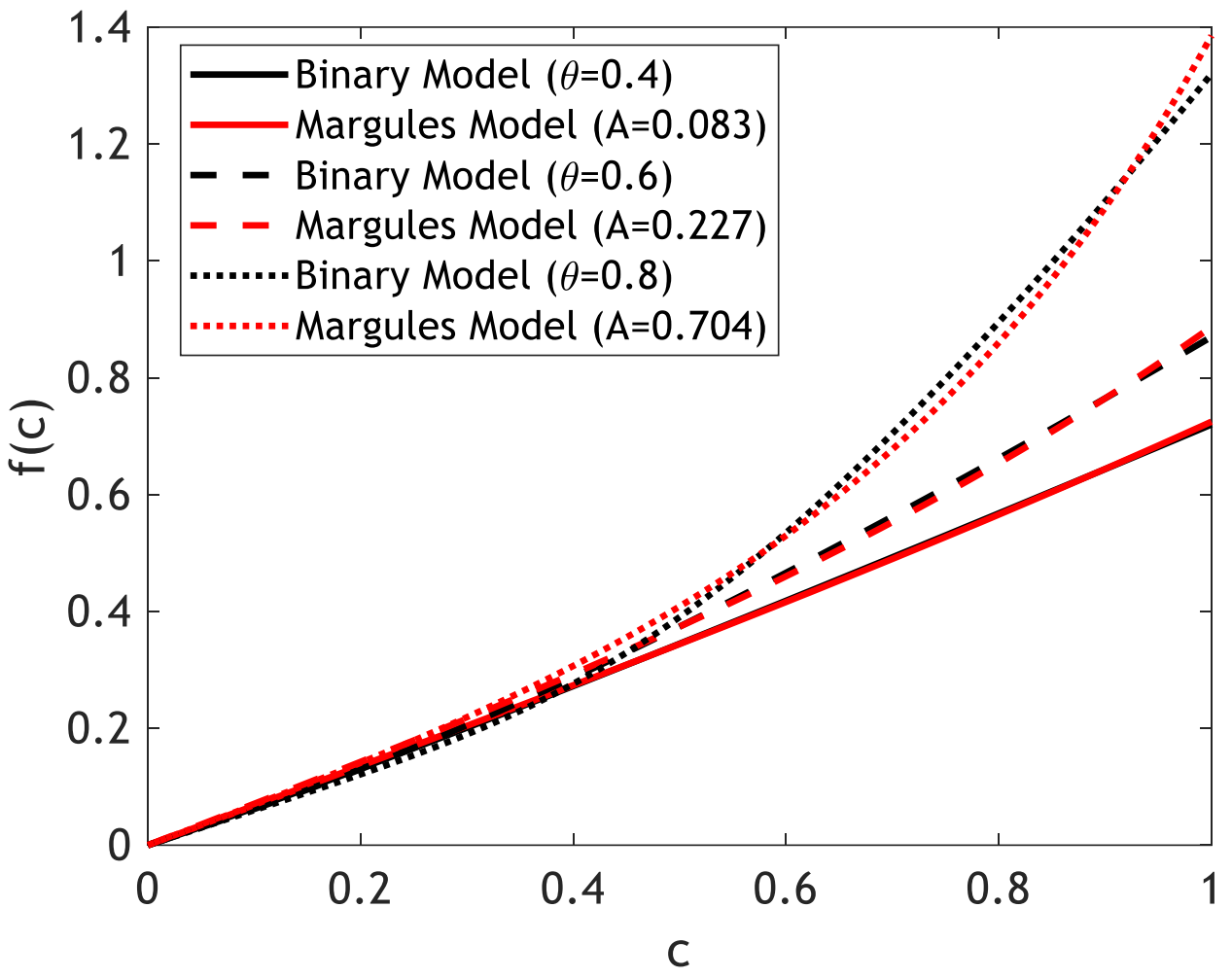

Fig. 3. Comparison between the binary and Margules models for the suspension function, *f(c)*

Table 1. Tuned values of $A$ obtained by matching the binary and Margules suspension functions (constant filtration function)

| $\theta$ \ $c_1^0$ | 0 | 0.1 | 0.2 | 0.3 | 0.4 | 0.5 | 0.6 | 0.7 | 0.8 | 0.9 | 1 |
|---|---|---|---|---|---|---|---|---|---|---|---|
| 1 | 0 | 9.8234 | 5.0850 | 3.3592 | 2.4075 | 1.7628 | 1.2168 | 0.8376 | 0.5128 | 0.2337 | 0.0000 |
| 0.9 | 0 | 0.8776 | 1.3283 | 1.3641 | 1.2213 | 1.0167 | 0.7957 | 0.5766 | 0.3675 | 0.1738 | 0.0016 |
| 0.8 | 0 | 0.3906 | 0.6211 | 0.7036 | 0.6870 | 0.6121 | 0.5051 | 0.3819 | 0.2526 | 0.1236 | 0.0012 |
| 0.7 | 0 | 0.2065 | 0.3352 | 0.3956 | 0.4037 | 0.3747 | 0.3207 | 0.2506 | 0.1707 | 0.0860 | 0.0009 |
| 0.6 | 0 | 0.1136 | 0.1876 | 0.2270 | 0.2383 | 0.2276 | 0.2002 | 0.1605 | 0.1121 | 0.0578 | 0.0006 |
| 0.5 | 0 | 0.0619 | 0.1038 | 0.1280 | 0.1372 | 0.1339 | 0.1203 | 0.0985 | 0.0702 | 0.0370 | 0.0004 |
| 0.4 | 0 | 0.0321 | 0.0545 | 0.0683 | 0.0745 | 0.0739 | 0.0676 | 0.0563 | 0.0409 | 0.0219 | 0.0002 |
| 0.3 | 0 | 0.0149 | 0.0257 | 0.0327 | 0.0361 | 0.0364 | 0.0338 | 0.0286 | 0.0210 | 0.0114 | 0.0001 |
| 0.2 | 0 | 0.0056 | 0.0098 | 0.0126 | 0.0140 | 0.0143 | 0.0135 | 0.0115 | 0.0086 | 0.0047 | $5.2\times10^{-5}$ |
| 0.1 | 0 | 0.0012 | 0.0021 | 0.0027 | 0.0031 | 0.0032 | 0.0030 | 0.0026 | 0.0020 | 0.0011 | $1.2\times10^{-5}$ |
| 0 | 0 | 0 | 0 | 0 | 0 | 0 | 0 | 0 | 0 | 0 | 0 |

Table 2. Tuned values of $\lambda$ obtained by matching the binary and Margules suspension functions (constant filtration function)

| $\theta$ \ $c_1^0$ | 0 | 0.1 | 0.2 | 0.3 | 0.4 | 0.5 | 0.6 | 0.7 | 0.8 | 0.9 | 1 |
|---|---|---|---|---|---|---|---|---|---|---|---|
| 1 | 0.6 | 0.0037 | 0.8657 | 7.3743 | 25.4626 | 60.0441 | 12.2199 | 20.2572 | 30.9734 | 44.3715 | 59.9118 |
| 0.9 | 0.6 | 0.4553 | 0.4623 | 0.6150 | 0.8966 | 1.3170 | 1.8944 | 2.6469 | 3.5861 | 4.7108 | 5.9864 |
| 0.8 | 0.6 | 0.5675 | 0.5959 | 0.6855 | 0.8332 | 1.0389 | 1.3046 | 1.6333 | 2.0265 | 2.4836 | 2.9946 |
| 0.7 | 0.6 | 0.6056 | 0.6418 | 0.7082 | 0.8043 | 0.9297 | 1.0845 | 1.2691 | 1.4836 | 1.7277 | 1.9971 |
| 0.6 | 0.6 | 0.6196 | 0.6553 | 0.7069 | 0.7742 | 0.8569 | 0.9551 | 1.0686 | 1.1975 | 1.3414 | 1.4983 |
| 0.5 | 0.6 | 0.6231 | 0.6546 | 0.6944 | 0.7425 | 0.7987 | 0.8631 | 0.9355 | 1.0158 | 1.1040 | 1.1990 |
| 0.4 | 0.6 | 0.6214 | 0.6471 | 0.6769 | 0.7109 | 0.7490 | 0.7912 | 0.8374 | 0.8876 | 0.9419 | 0.9994 |
| 0.3 | 0.6 | 0.6173 | 0.6365 | 0.6575 | 0.6805 | 0.7053 | 0.7320 | 0.7606 | 0.7909 | 0.8231 | 0.8568 |
| 0.2 | 0.6 | 0.6119 | 0.6245 | 0.6378 | 0.6518 | 0.6664 | 0.6818 | 0.6978 | 0.7145 | 0.7319 | 0.7498 |
| 0.1 | 0.6 | 0.6060 | 0.6122 | 0.6185 | 0.6249 | 0.6315 | 0.6383 | 0.6452 | 0.6667 | 0.6594 | 0.6666 |
| 0 | 0.6 | 0.6 | 0.6 | 0.6 | 0.6 | 0.6 | 0.6 | 0.6 | 0.6 | 0.6 | 0.6 |

Table 3. Obtained values of $R^2$ obtained by matching the binary and Margules suspension functions (constant filtration function)

| $\theta$ \ $c_1^0$ | 0 | 0.1 | 0.2 | 0.3 | 0.4 | 0.5 | 0.6 | 0.7 | 0.8 | 0.9 | 1 |
|---|---|---|---|---|---|---|---|---|---|---|---|
| 1 | 1 | 0.9665 | 0.9572 | 0.9492 | 0.9434 | 0.9417 | 0.9491 | 0.958 | 0.9727 | 0.9899 | 1 |
| 0.9 | 1 | 0.9928 | 0.9973 | 0.991 | 0.982 | 0.976 | 0.9748 | 0.9785 | 0.986 | 0.995 | 1 |
| 0.8 | 1 | 0.9998 | 0.9992 | 0.9961 | 0.9921 | 0.9892 | 0.9884 | 0.99 | 0.9936 | 0.9977 | 1 |
| 0.7 | 1 | 1 | 0.9993 | 0.9979 | 0.9963 | 0.9951 | 0.9948 | 0.9956 | 0.9972 | 0.999 | 1 |
| 0.6 | 1 | 0.9999 | 0.9996 | 0.999 | 0.9983 | 0.9979 | 0.9979 | 0.9982 | 0.9989 | 0.9996 | 1 |
| 0.5 | 1 | 0.9999 | 0.9998 | 0.9996 | 0.9993 | 0.9992 | 0.9992 | 0.9993 | 0.9996 | 0.9999 | 1 |
| 0.4 | 1 | 1 | 0.9999 | 0.9998 | 0.9998 | 0.9997 | 0.9997 | 0.9998 | 0.9999 | 1 | 1 |
| 0.3 | 1 | 1 | 1 | 1 | 0.9999 | 0.9999 | 0.9999 | 0.9999 | 1 | 1 | 1 |
| 0.2 | 1 | 1 | 1 | 1 | 1 | 1 | 1 | 1 | 1 | 1 | 1 |
| 0.1 | 1 | 1 | 1 | 1 | 1 | 1 | 1 | 1 | 1 | 1 | 1 |
| 0 | 1 | 1 | 1 | 1 | 1 | 1 | 1 | 1 | 1 | 1 | 1 |

The values of $A$ in Table 1 are highest when the difference between the two filtration coefficients of the binary model is highest ($\theta$~1), and when $c_1^0$ is close to zero, indicating that the Margules model which captures the behaviour of the binary model exhibits the highest non-linearity when the fraction of particles with high filtration coefficient is smaller than that with lower filtration coefficient. On the other hand, in Table 2 we see that the single filtration coefficient for the Margules model is highest around ($\theta$, $c_1^0$)=(1,1), which is expected, given that these conditions correspond to a high fraction of high filtration coefficient particles.

The coefficient of determination values in Table 3 show good agreement between the two models across all values of the binary model coefficients. The binary model exhibits the greatest non-linearity in the suspension function when neither of the two particle populations dominate ($c_1^0 \sim 0.5$) and one filtration coefficient greatly exceeds the other ($\theta \sim 1$). Precisely under these conditions we observe a decrease in the quality of agreement between the two models. While the tuned values of $\lambda$ reach a maximum in this region, the highest values of $A$ appear closer to $c_1^0 = 0$.

To facilitate the substitution of one model by another, the values of $A$ presented in Table 1 are tuned by a polynomial to allow for an explicit expression:

$$\begin{aligned}
A(c_1^0,\theta) = {} & P_{00} + P_{10}c_1^0 + P_{01}\theta + P_{20}\left(c_1^0\right)^2 + P_{11}c_1^0\theta + P_{02}\theta^2 + P_{30}\left(c_1^0\right)^3 + P_{21}\left(c_1^0\right)^2\theta + P_{12}c_1^0\theta^2 + \\
& P_{03}\theta^3 + P_{40}\left(c_1^0\right)^4 + P_{31}\left(c_1^0\right)^3\theta + P_{22}\left(c_1^0\right)^2\theta^2 + P_{13}c_1^0\theta^3 + P_{04}\theta^4 + P_{50}\left(c_1^0\right)^5 + P_{41}\left(c_1^0\right)^4\theta + P_{32}\left(c_1^0\right)^3\theta^2 \\
& P_{23}\left(c_1^0\right)^2\theta^3 + P_{14}c_1^0\theta^4 + P_{05}\theta^5
\end{aligned} \quad (32)$$

This matching had a coefficient of determination, $R^2$ of 0.7765. Contours of this polynomial are shown in Fig. 4. This figure shows the concentration of high values of $A$ around high values of $\theta$, indicating that non-linearity in the Margules model corresponds to the case when the two particles in the binary model exhibit greatly varying capture properties. Two additional contours for $A$=0.01 are found near $\theta$=0, indicating a non-monotonic trend of $A$. This is not present in Table 1 and thus is purely an artifact of the tuning of the values in Table 1 with Eq. (32). Calculation of $A$ values in this region using the method of performing the minimisation in Eq. (31) would likely result in values of $A$ in these regions below 0.01.

The values of the coefficients for Eq. (32) are presented in Table 4.

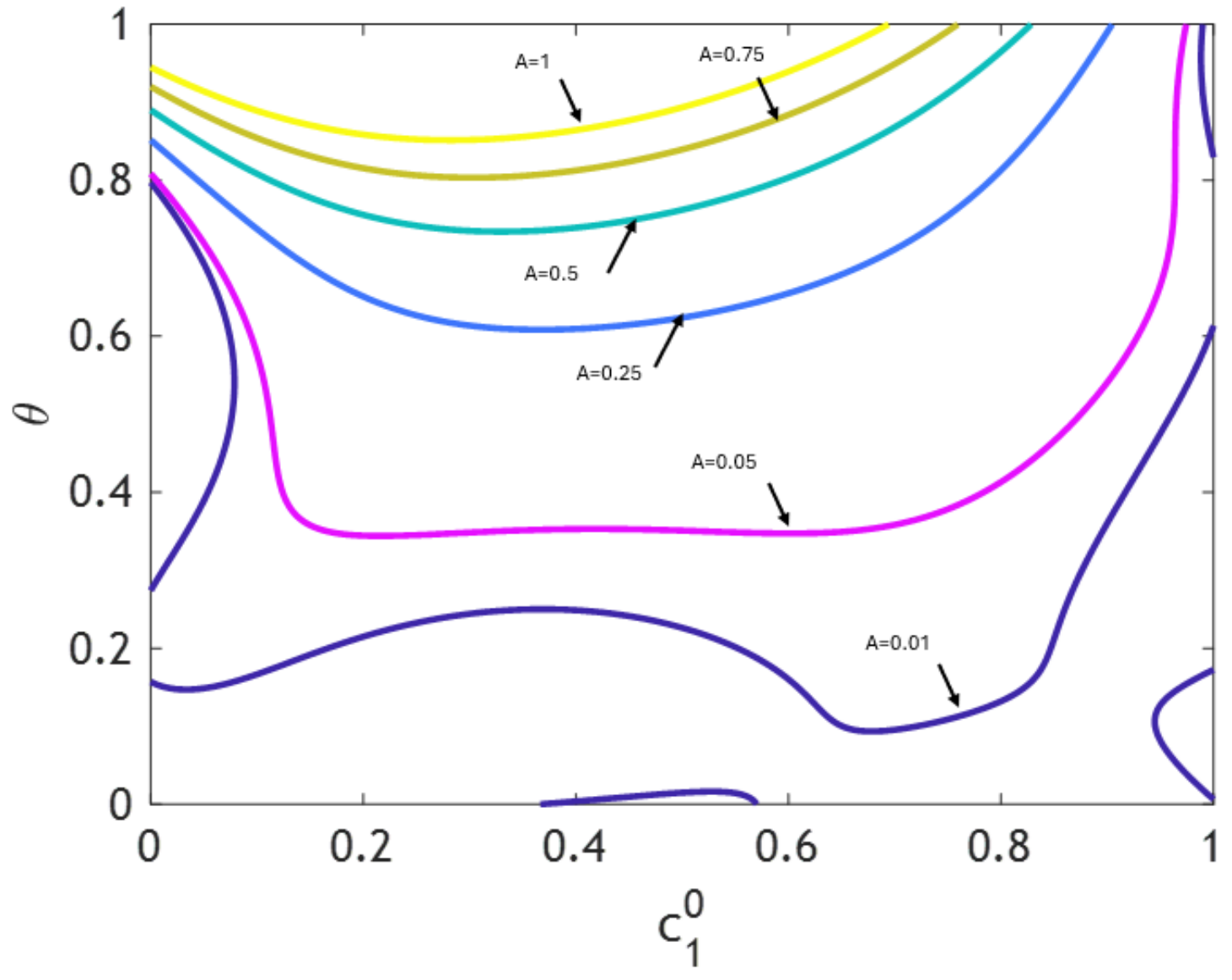

Fig. 4. Contours of polynomial (32) for $A$ obtained by matching the binary and Margules suspension functions (constant filtration function)

Table 4. Coefficients of the polynomial fit for $A$ obtained by matching the binary and Margules suspension functions (constant filtration function)

| Coefficient | Value | Coefficient | Value |
|---|---|---|---|
| $P_{00}$ | -0.0094 | $P_{41}$ | -6.6253 |
| $P_{10}$ | 0.2042 | $P_{14}$ | -15.889 |
| $P_{01}$ | -0.0356 | $P_{42}$ | 0 |
| $P_{11}$ | -2.1168 | $P_{24}$ | 0 |
| $P_{20}$ | -1.296 | $P_{43}$ | 0 |
| $P_{02}$ | 1.9717 | $P_{34}$ | 0 |
| $P_{21}$ | 1.2678 | $P_{44}$ | 0 |
| $P_{12}$ | 6.9748 | $P_{50}$ | 2.762 |
| $P_{22}$ | -27.232 | $P_{05}$ | 5.7541 |
| $P_{30}$ | 4.1757 | $P_{51}$ | 0 |
| $P_{03}$ | -6.3007 | $P_{15}$ | 0 |
| $P_{31}$ | 8.0892 | $P_{52}$ | 0 |
| $P_{13}$ | 18.441 | $P_{25}$ | 0 |
| $P_{32}$ | 12.971 | $P_{53}$ | 0 |
| $P_{23}$ | 2.3771 | $P_{35}$ | 0 |
| $P_{33}$ | 0 | $P_{54}$ | 0 |
| $P_{40}$ | -5.8299 | $P_{45}$ | 0 |
| $P_{04}$ | 0.3305 | $P_{55}$ | 0 |

Comparison between the two models can also be performed by comparing the output measurements of a typical coreflooding experiment. In this way the optimal value of $A$ and $\lambda$ corresponds to the minimisation of quantities that may be desired when performing predictive calculations. Any combination of measurements could be chosen here, which may alter slightly the value of $(A, \lambda)$ obtained through optimization. In this study, we choose the outlet suspended particle concentration (versus time), and the retained particle concentration profile (versus distance) at a particular time. Thus, the optimization corresponds to the following minimisation:

$$\min_{(A,\lambda)}\left[\left(c_B(1,t)-c_M(1,t,A,\lambda)\right)^2+\left(s_B(x,t^*)-s_M(x,t^*,A,\lambda)\right)^2\right] \quad (33)$$

Where $c_B(c)$ and $c_M(c)$ are the binary and Margules suspended particle concentrations, and $s_B(c)$ and $s_M(c)$ are the binary and Margules retained particle concentrations respectively.

The results of this analysis are similar using both methods. Corresponding tables and contour plot are presented in the supplementary material for the case of using the BTC and RP. Tuning of the $A$ values using a polynomial resulted in a coefficient of determination, $R^2$, of 0.7765 and 0.8813 for the case of tuning $f(c)$ and the BTC/RP methods respectively. The contour plot for $A$ presented in Fig. S1 shows a similar trend of $A(\theta, c_1^0)$ except for a more symmetric dependence of $A$ on $c_1^0$. This suggests that when considering only the suspension function predicts more non-linearity (higher $A$) at lower values of $c_1^0$, whereas when treating the output of a laboratory test, no such preference exists.

## 5. Matching laboratory coreflooding data

This section presents results of treatment of the laboratory data using the traditional model ($f(c)=\lambda c$) and the general model (7),(8) with binary and Margules suspension functions; the corresponding analytical models are given by Eqs. (28),(29) (classical) and Eqs. (24),(25),(26), respectively. Section 5.1 matches the laboratory data on pressure drop with varying flow rates and injected concentrations. Section 5.2 matches the tests with BTCs and RPs for different values of solute ionic strength and pH.

### 5.1. Matching pressure drop data

Ramachandran and Fogler [43], [44] injected charged polystyrene microspheres dispersed in water into multi-layer porous membranes. A significant number of overlapping pores and strong inter-pore connections allowed claiming the porous space structure as an irregular pore network. Zeta-potential measurements and DLVO calculations revealed strong electrostatic 'particle-particle' and 'particle-matrix' repulsion, so the capture mechanisms by aggregation and consequent aggregate straining were excluded. Size exclusion and straining were also excluded under the pores radii that highly exceed the particle radii. However, the experiments show substantial particle retention. The established mechanism causing particle retention under the test conditions was a flow-induced hydrodynamic bridging.

The experimental data matching by three models are performed using nonlinear least square method, which minimises the normalised deviation between the modelling and the experimental data [45]. The so-called Trust-Region-Reflective optimisation algorithm solves the problem of quadratic-deviation minimisation. This algorithm is implemented in Matlab. Besides high agreement with the experimental data, the model validation must include some predictive modelling, i.e. the number of tuned model coefficients must be smaller than the dimension (number of degrees of freedom) of the matched data array [46].

Figs. 5 and 6 present pressure drop measurements for three series per three tests each, with injection of particles with different rates. The points $(J(t_k)-1)/t_k$, $k$=1, 2… as calculated from the laboratory data do not fit to straight lines for all the tests. In Fig. 5, we tune all three curves simultaneously assuming, where relevant, that parameters ($A$,$\beta$, $c_1^0$,$s_m$) are constants of the system and do not vary with velocity. The values of lambdas are assumed to be functions of velocity and thus the values for each curve are matched independently.

The results presented in Fig. 5 and Table 5 show that the classical model is unable to capture the rapidly increasing impedance resulting from particle capture. Both the binary and Margules model capture the trend well, indicating that they are correctly capturing the growth of the retention profile. In addition, matching of the three tests exhibit filtration coefficient that monotonically increases with velocity (Table 5), which agrees with the classical filtration theory [13].

Tuning in Fig. 6 is performed using only two curves, under the assumption that each curve contains three degrees of freedom, enough to tune the unknown parameters of each model. The third curve at the lowest velocity is tuned separately, tuning only the filtration coefficient (see Table 6). This procedure serves as a more rigorous test of the agreement between the models and the experimental data. The high quality of agreement for the binary and Margules models again confirms that their non-linear suspension functions can capture the underlying filtration behaviour of these tests.

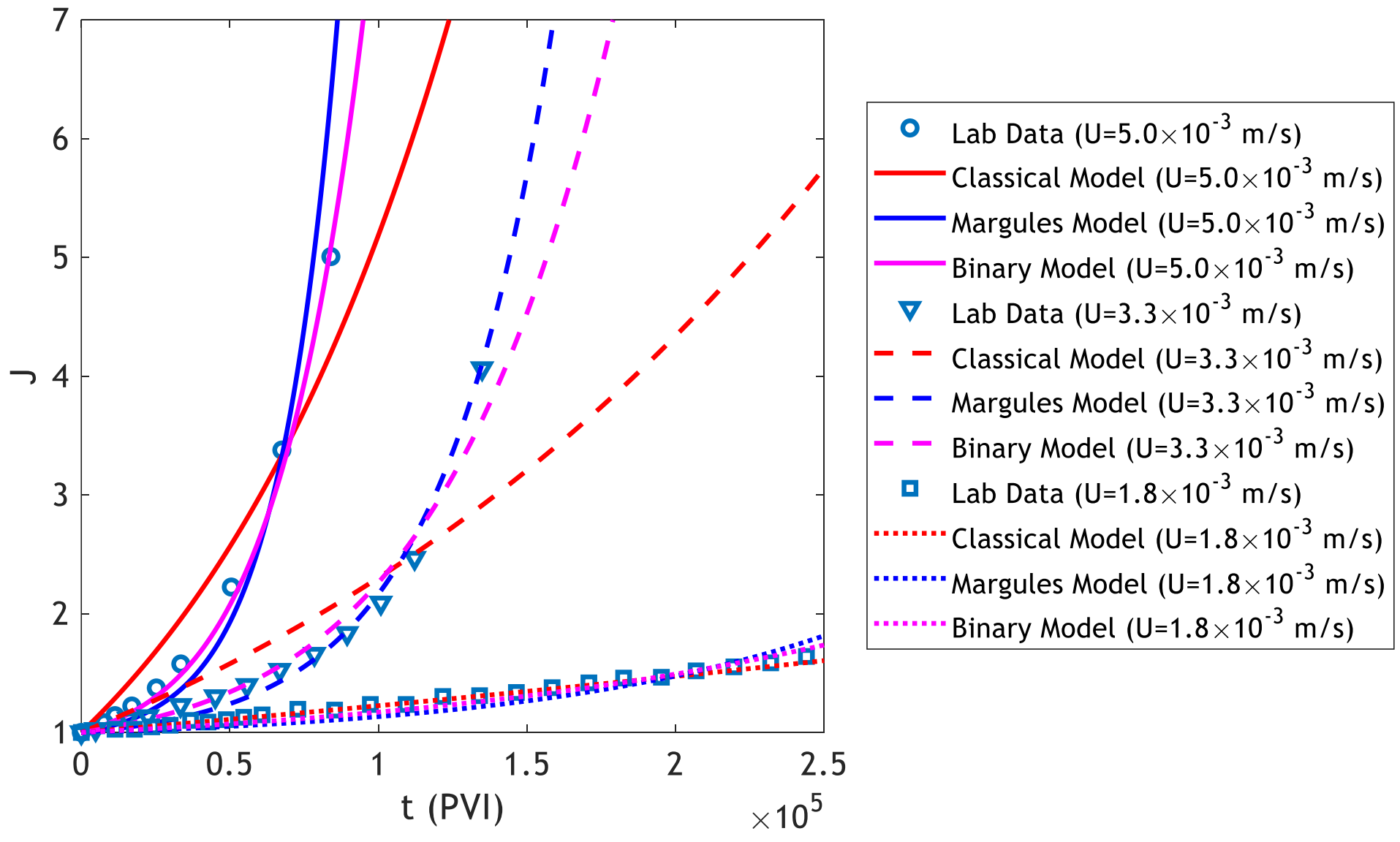

Fig. 5. Treatment of impedance lab data taken from Ramachandran and Fogler [43]

Table 5. Tuning parameters for treatment of impedance lab data taken from Ramachandran and Fogler [43] (Fig. 5)

| Models | Test Conditions | $A$ ×10$^{-5}$ | $\beta$ | $c_1^0$ [ppm] | $s_m$ [L$^{-3}$] | $\lambda_1$ [m$^{-1}$] | $\lambda_2$ [m$^{-1}$] | $R_p^2$ | Dataset |
|---|---|---|---|---|---|---|---|---|---|
| | $c^0$= 1235 ppm U= 0.0050 m/s | | | | | 3.86×10$^{-2}$ | | 0.924 | Tuning |
| Classical | $c^0$= 1235 ppm U= .0.0033 m/s | | 5.02E×10$^5$ | | -4.61×10$^{-6}$ | 1.67×10$^{-2}$ | | 0.788 | Tuning |
| | $c^0$= 1235 ppm U= 0.0018 m/s | | | | | 3.47×10$^{-3}$ | | 0.989 | Tuning |
| | $c^0$= 1235 ppm U= 0.0050 m/s | | | | | 1.58×10$^{-4}$ | | 0.863 | Tuning |
| Margules | $c^0$= 1235 ppm U= 0.0033 m/s | 13.494 | 2.77×10$^6$ | | -3.08×10$^{-8}$ | 8.55×10$^{-5}$ | | 0.988 | Tuning |
| | $c^0$= 1235 ppm U= 0.0018 m/s | | | | | 3.00×10$^{-5}$ | | 0.854 | Tuning |
| | $c^0$= 1235 ppm U= 0.0050 m/s | | | | | 4.96×10$^{-2}$ | 6.00×10$^{-3}$ | 0.995 | Tuning |
| Binary | $c^0$= 1235 ppm U= 0.0033 m/s | | 1.07×10$^6$ | 459.42 | -1.52×10$^{-7}$ | 4.05×10$^{-2}$ | 5.10×10$^{-3}$ | 0.966 | Tuning |
| | $c^0$= 1235 ppm U= 0.0018 m/s | | | | | 2.00×10$^{-2}$ | 4.90×10$^{-3}$ | 0.946 | Tuning |

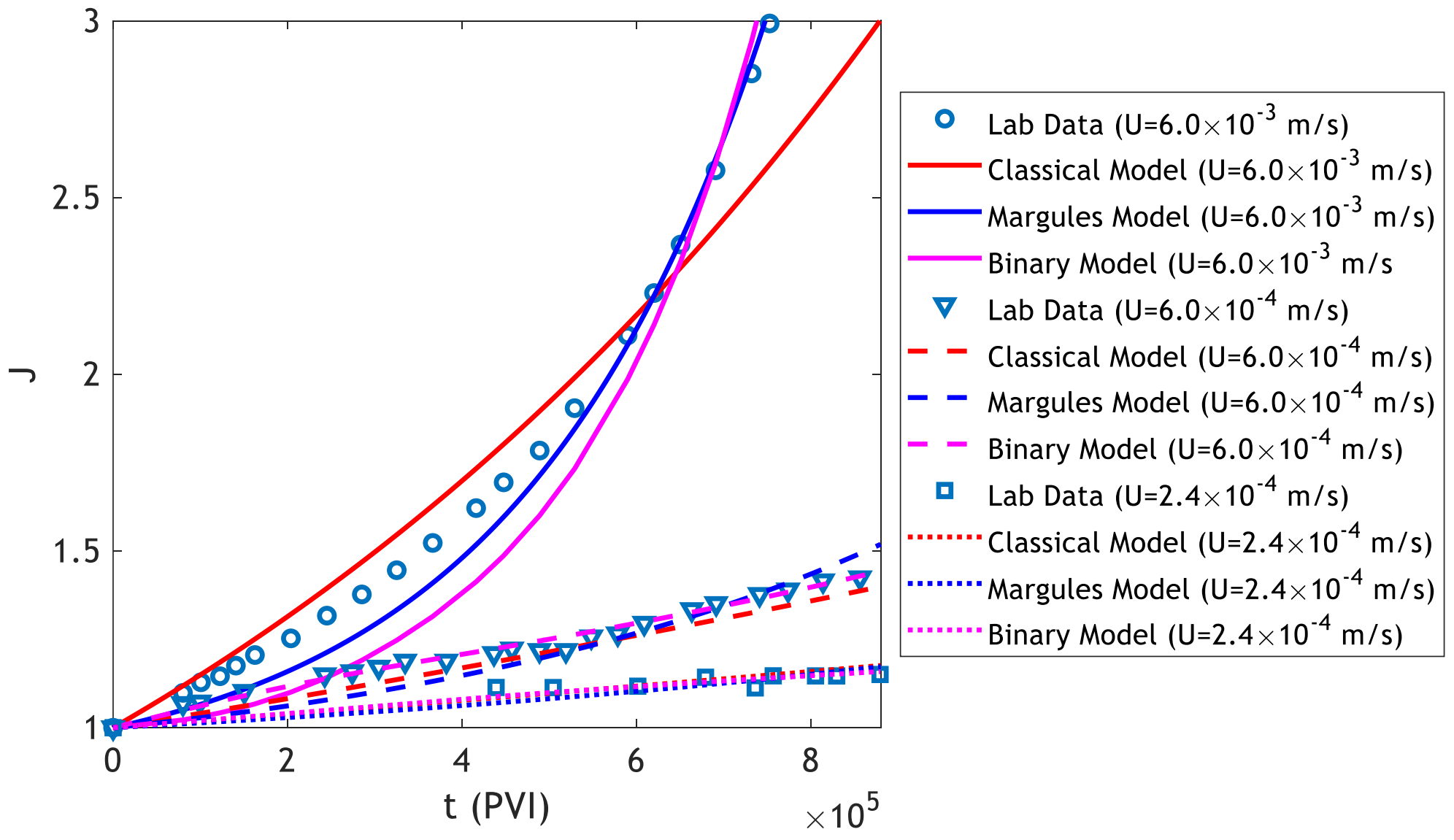

Fig. 6. Treatment of impedance lab data taken from Ramachandran and Fogler [43]

Table 6. Tuning parameters for treatment of impedance lab data taken from Ramachandran and Fogler [43] (Fig. 6)

<table>
<tr><th>Models</th><th>Test Conditions</th><th>$A \times 10^{-5}$</th><th>$\beta$</th><th>$c_1^0$ [ppm]</th><th>$s_m$ [$L^{-3}$]</th><th>$\lambda_1$ [$m^{-1}$]</th><th>$\lambda_2$ [$m^{-1}$]</th><th>$R_p^2$</th><th>Dataset</th></tr>
<tr><td rowspan="3">Classical</td><td>$c^0$= 415 ppm<br>U= 0.006 m/s</td><td rowspan="3"></td><td rowspan="3">$1.04\times10^{5}$</td><td rowspan="3"></td><td rowspan="3">$-1.37\times10^{-5}$</td><td>$3.30\times10^{-2}$</td><td></td><td>0.947</td><td>Tuning</td></tr>
<tr><td>$c^0$= 415 ppm<br>U=0.0006 m/s</td><td>$9.28\times10^{-3}$</td><td></td><td>0.902</td><td>Tuning</td></tr>
<tr><td>$c^0$= 415 ppm<br>U=0.00024 m/s</td><td>$4.38\times10^{-3}$</td><td></td><td>0.811</td><td>Prediction</td></tr>
<tr><td rowspan="3">Margules</td><td>$c^0$= 415 ppm<br>U= 0.006 m/s</td><td rowspan="3">125.603</td><td rowspan="3">$1.71\times10^{6}$</td><td rowspan="3"></td><td rowspan="3">$-9.17\times10^{-8}$</td><td>$8.91\times10^{-5}$</td><td></td><td>0.985</td><td>Tuning</td></tr>
<tr><td>$c^0$= 415 ppm<br>U=0.0006 m/s</td><td>$4.22\times10^{-5}$</td><td></td><td>0.855</td><td>Tuning</td></tr>
<tr><td>$c^0$= 415 ppm<br>U=0.00024 m/s</td><td>$2.14\times10^{-5}$</td><td></td><td>0.741</td><td>Prediction</td></tr>
<tr><td rowspan="3">Binary</td><td>$c^0$= 415 ppm<br>U= 0.006 m/s</td><td rowspan="3"></td><td rowspan="3">$1.28\times10^{6}$</td><td rowspan="3">14.140</td><td rowspan="3">$-1.54\times10^{-7}$</td><td>$1.98\times10^{1}$</td><td>6.60</td><td>0.940</td><td>Tuning</td></tr>
<tr><td>$c^0$= 415 ppm<br>U=0.0006 m/s</td><td>1.33</td><td>$1.33\times10^{-2}$</td><td>0.984</td><td>Tuning</td></tr>
<tr><td>$c^0$= 415 ppm<br>U=0.00024 m/s</td><td>1.18</td><td>$1.19\times10^{-2}$</td><td>0.900</td><td>Prediction</td></tr>
</table>

## 5.2. Matching suspension and retained concentrations

The large-scale models (7),(8) are diffusion / dispersion free, yielding suspension concentration discontinuity across the front, while the concentration jumps in BTCs are smoothed by micro-scale dissipative processes. Therefore, we treat the steep continuous growth of a BTC around its breakthrough time at the moment of one pore volume injected (PVI) as a concentration jump. Suspension concentration before the breakthrough is zero. The jump occurs to continuously growing BTC after the breakthrough. The areas between the continuous and discontinuous BTCs before and after the breakthrough time are equal. This BTC-treatment procedure is analogous to matching of inner and outer asymptotic expansions in the case of small diffusion and dispersion [46-50].

Fig. 7 shows a set of laboratory data where different salinities were injected [25]. Spherical soda lime glass beads have been used for packing the column with length 20 cm and diameter 3.81 cm. Spherical fluorescent carboxylate-modified polystyrene latex particles with diameter 1.1 μm were used. For each test, filtration coefficient, maximum retained concentration, and $A$-value are determined by simultaneous matching BTC and RP. As with the previous test, we see good agreement for the binary and Margules models, while the classical model fails to match the hyper-exponential retention profiles. For both cases, the higher is the salinity the higher is the filtration coefficient (see Table 7); this is explained by increase in electrostatic attraction as the solute ionic strength increases.

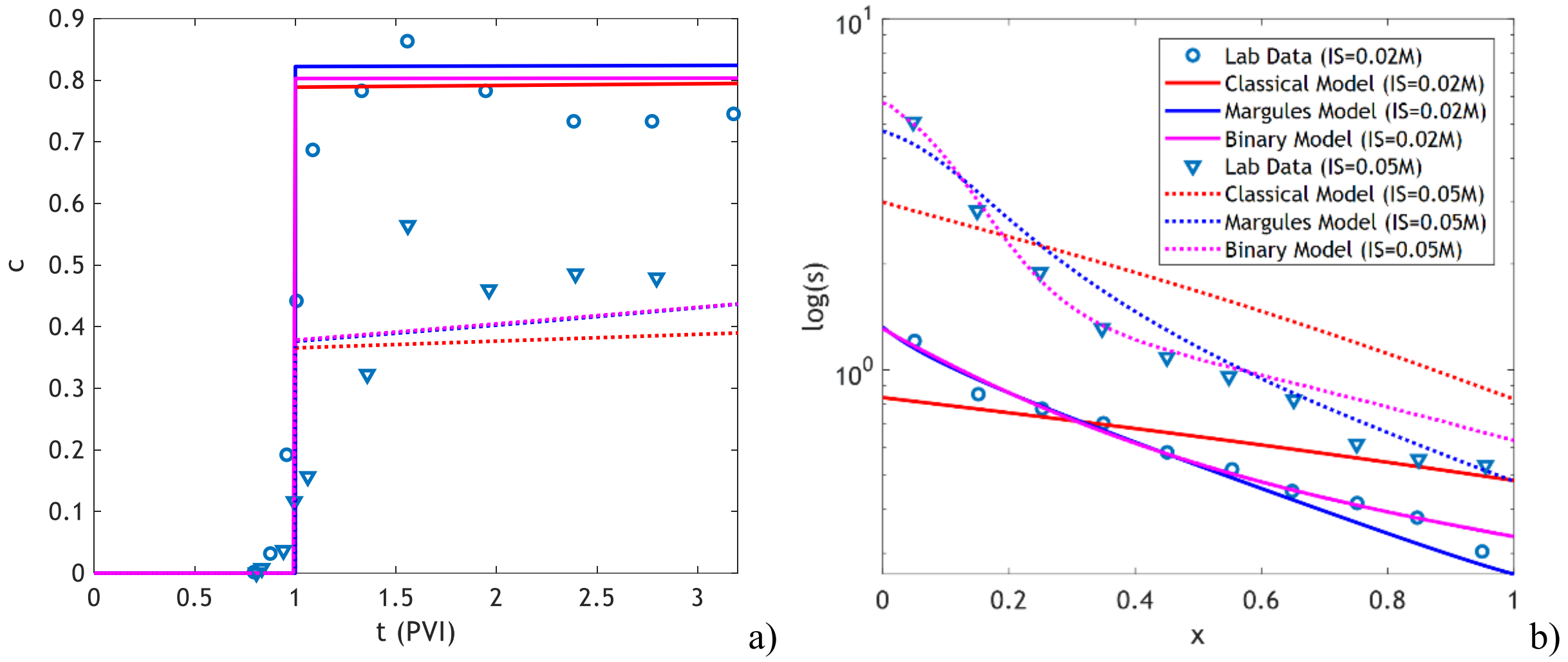

Fig. 7. Simultaneous treatment of the suspended and retained concentrations of lab data taken from Li, Scheibe and Johnson [25]

Table 7. Tuning parameters for treatment of breakthrough concentration and retention profiles data taken from Li, Scheibe and Johnson [25] (Fig. 7)

| Models | Conditions of the Test | $A \times 10^{-5}$ | $c_1^0$ [ppm] | $s_m$ [L$^{-3}$] | $\lambda_1$ [m$^{-1}$] | $\lambda_2$ [m$^{-1}$] | $R_c^2$ | $R_s^2$ |
|---|---|---|---|---|---|---|---|---|
| Classical | $c^0$= 193 ppm IS= 0.02 M | | | $1.02\times10^{-3}$ | 1.917 | | 0.832 | 0.576 |
| Margules | | $9.40\times10^{2}$ | | $1.46\times10^{-3}$ | 0.092 | | 0.789 | 0.970 |
| Binary | | | 28.986 | $7.20\times10^{-3}$ | 26.368 | 1.216 | 0.817 | 0.979 |
| Classical | $c^0$= 193 ppm IS= 0.05 M | | | $1.55\times10^{-3}$ | 5.032 | | 0.727 | 0.585 |
| Margules | | $7.08\times10^{2}$ | | $3.57\times10^{-4}$ | 1.821 | | 0.753 | 0.947 |
| Binary | | | 40.535 | $5.00\times10^{-4}$ | 84.647 | 3.680 | 0.775 | 0.991 |

Figs. 8 and 9 presents the results of colloidal injection into porous column packed by quartz sand with 425-600 μm [51]. The column length was 12 cm and the diameter was 2.5 cm. BTCs and RPs with injection of biochar colloids are shown in Figs. 8 for three different pH values. The curves presented in Fig. 9 correspond to biochar colloids coated by Cytochrome protein BTC shows the tendency of low decrease, so the model with blocking filtration function (19, 20) was used. Three constants $\lambda$, $s_m$, and $A$ were determined from each test. The data are for Figs. 8 and 9 are presented in Tables 8 and 9 respectively.

The results show a similar trend. In all cases all three models show good agreement with the breakthrough concentration, largely because the BTC, measured only over several pore volumes, is approximately constant. The retention profiles in the *log(s)* vs *x* plots show non-linear trends, indicating hyper-exponentiality. This is captured by both the binary and Margules models, but not by the classical model. For both the coated and uncoated biochar colloids, lower pH leads to higher values of the filtration coefficient, consistent with the DLVO theory of electrostatic attraction between particles and porous media [52].

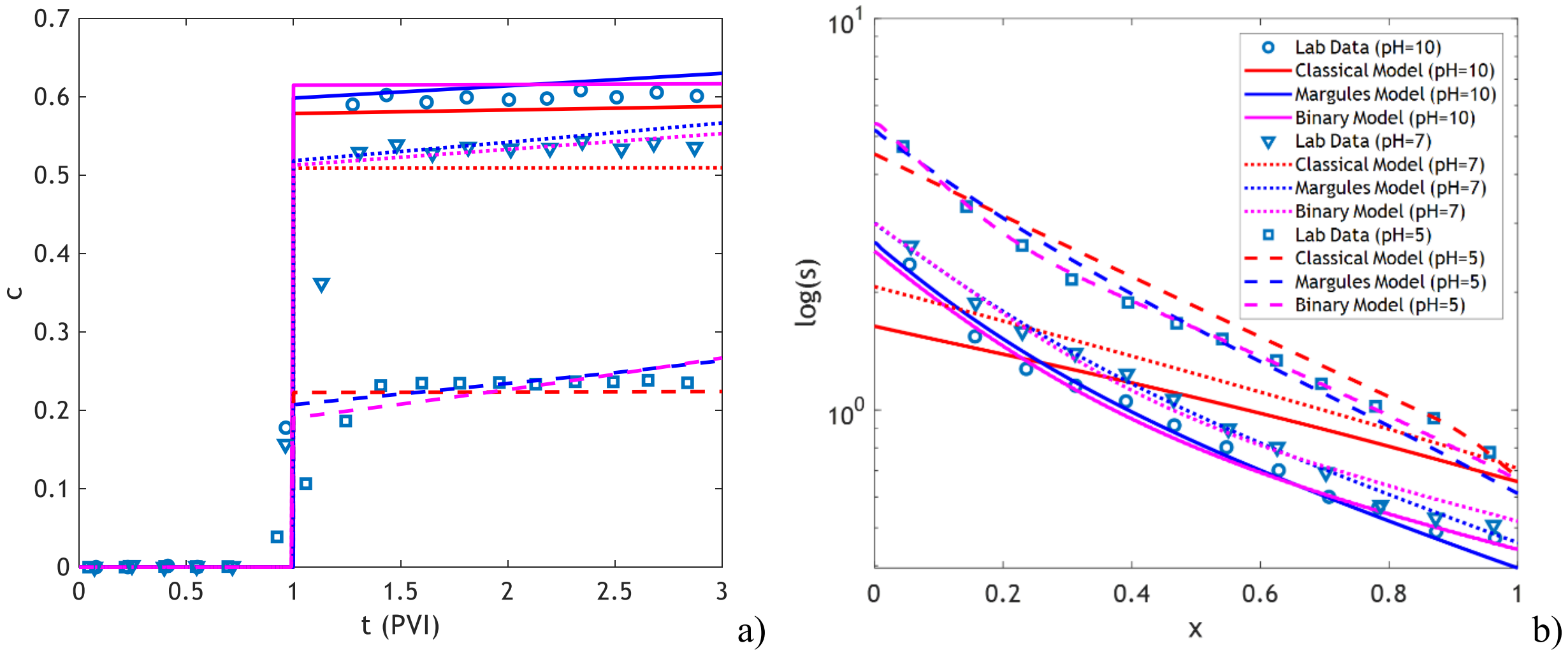

Fig. 8. Simultaneous treatment of the suspended and retained concentrations of lab data taken from Yang, Bradford, Wang, Sharma, Shang and Li [51]

Table 8. Tuning parameters for treatment of breakthrough concentration and retention profiles data taken from Yang, Bradford, Wang, Sharma, Shang and Li [51] (Fig. 8)

| Models | Conditions of the Test | $A \times 10^{-5}$ | $c_1^0$ [ppm] | $s_m$ [$L^{-3}$] | $\lambda_1$ [$m^{-1}$] | $\lambda_2$ [$m^{-1}$] | $R_c^2$ | $R_s^2$ |
|---|---|---|---|---|---|---|---|---|
| Classical | $c^0$= 50 ppm<br>IS = 0.001 M<br>BCs, pH=10 | | | $6.85\times10^{-4}$ | 4.563 | | 0.969 | 0.668 |
| Margules | | 8802.8 | | $1.14\times10^{-4}$ | 1.170 | | 0.969 | 0.982 |
| Binary | | | 6.69 | $2.35\times10^{-3}$ | 34.996 | 2.88 | 0.969 | 0.979 |
| Classical | $c^0$= 50 ppm<br>IS = 0.001 M<br>BCs, pH=7 | | | $1.61\times10^{-2}$ | 5.630 | | 0.948 | 0.767 |
| Margules | | 7524 | | $1.04\times10^{-4}$ | 2.063 | | 0.948 | 0.994 |
| Binary | | | 7.63 | $1.47\times10^{-4}$ | 50.803 | 4.189 | 0.952 | 0.99 |
| Classical | $c^0$= 50 ppm<br>IS = 0.001 M<br>BCs, pH=5 | | | $1.08\times10^{-2}$ | 12.642 | | 0.911 | 0.938 |
| Margules | | 2843.6 | | $2.03\times10^{-4}$ | 10.737 | | 0.927 | 0.981 |
| Binary | | | 5.875 | $1.65\times10^{-4}$ | 146.879 | 12.772 | 0.939 | 0.998 |

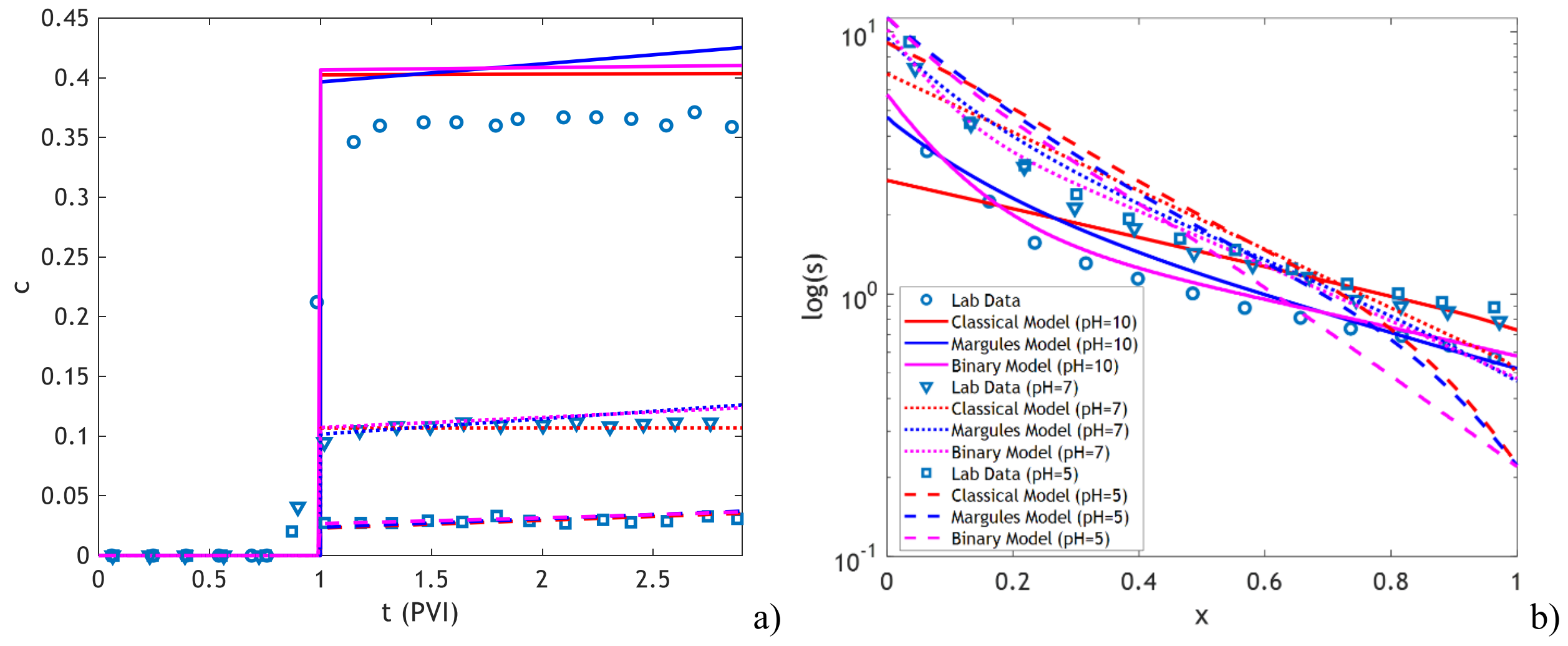

Fig. 9. Simultaneous treatment of the suspended and retained concentrations of lab data taken from Yang, Bradford, Wang, Sharma, Shang and Li [51]

Table 9. Tuning parameters for treatment of breakthrough concentration and retention profiles data taken from Yang, Bradford, Wang, Sharma, Shang and Li [51] (Fig. 9)

| Models | Conditions of the Test | $A \times 10^{-5}$ | $c_1^0$ [ppm] | $s_m$ [$L^{-3}$] | $\lambda_1$ [1/m] | $\lambda_2$ [1/m] | $R_c^2$ | $R_s^2$ |
|---|---|---|---|---|---|---|---|---|
| Classical | $c^0$= 50 ppm | | | $8.78\times10^{-3}$ | 7.475 | | 0.877 | 0.682 |
| Margules | IS = 0.001 M | 5617.2 | | $2.45\times10^{-4}$ | 4.282 | | 0.855 | 0.903 |
| Binary | Cyt, pH=10 | | 6.95 | $2.20\times10^{-3}$ | 84.215 | 6.25 | 0.865 | 0.973 |
| Classical | $c^0$= 50 ppm | | | $1.04\times10^{-1}$ | 18.646 | | 0.957 | 0.893 |
| Margules | IS = 0.001 M | 3389.6 | | $3.87\times10^{-4}$ | 16.180 | | 0.933 | 0.948 |
| Binary | Cyt, pH=7 | | 7.27 | $5.73\times10^{-4}$ | 149.043 | 17.2883 | 0.936 | 0.983 |
| Classical | $c^0$= 50 ppm | | | $5.57\times10^{-4}$ | 31.448 | | 0.814 | 0.83 |
| Margules | IS = 0.001 M | 1628.4 | | $4.92\times10^{-4}$ | 29.636 | | 0.790 | 0.868 |
| Binary | Cyt, pH=5 | | 6.33 | $4.86\times10^{-4}$ | 152.975 | 29.02 | 0.804 | 0.906 |

Three additional sets of experimental data from this study are tuned and presented in the supplementary material.

For all tests, the tuned model coefficients $\lambda$, $s_m$, and $A$ vary in the common intervals [2, 4, 7, 13, 17-19, 21, 26, 33, 53-59]. The $A$-data presented in the above Tables can be used for predictive modelling.

## 6. Discussions

The motivation of this work stems from failure of the traditional DBF equations (1, 3, 5) to match simultaneously BTC and PR for frequently encountered cases of HERPs. Therefore, the main aim of the work is a simplest possible extension of the traditional model to reproduce HERPs. The present paper transforms the traditional model to Eqs. (1, 3, 6) by introducing a new phenomenological suspension function *f(c)*. The extended model exhibits HERPs. The numerous laboratory data sets can be closely fitted and predicted using suspension function $f(c)=\lambda c\exp(Ac^2)$, which adds just one extra parameter into the traditional model.

*Validity of the assumptions and of the extended model* Failure of the traditional model to match HERPs is usually attributed to distributed properties of the suspended particles [8, 22, 31, 32]. We show that a non-linear suspension function can be obtained by averaging of the suspension-colloidal flow with distributed filtration coefficient, which in turn can be attributed to distributed size, shape, surface charge and other particle properties (Appendix C). Therefore, the proposed model implicitly accounts for the particle property distribution. Moreover, it has been shown that the wider is the continuous filtration coefficient distribution the higher is the second derivative of suspension function [40], and the higher is the deviation between the RPs predicted by the traditional and modified models. The same applies to a binary colloid – the higher is the ratio between the filtration coefficients of two populations, the higher is the deviation between the two models. This averaging strongly supports the modified model and explains HERPs by heterogeneity of the colloidal flux.

The experimental verification of the above claim would include injection of binary colloids with various properties in identical porous columns. To the best of our knowledge, those laboratory data are not available in the literature.

We propose an alternative reason for deviation between the laboratory data and traditional modelling that is high suspension concentration, which promotes the competition between the suspended particles for the vacancies. At high suspension concentrations, where the particles compete for vacancies, the events of particle appearance near the vacancies are not independent, so Eq. (2) is not valid.

The use of Margules' model for activity coefficients to treat the problem of deep bed filtration is justified by an analogy between chemical reactions and filtration events. Just as chemical reactions rely on the interception between different molecules, filtration occurs when a suspended particle (or multiple) interacts with a capture site. Under ideal conditions, the reaction kinetics are proportional to concentrations, just as the classical model for deep bed filtration is proportional to the suspended particle concentration (absence of the concentration of capture sites is due to the common assumption that they greatly outnumber the retained concentration). Similar to the breakdown of the 'ideal' solution assumption for chemical reactions, the results of this study show that filtration rates vary nonlinearly with species concentration. This study only considers a small number of experimental studies so it is possible for other experiments to justify the use of other activity models such as the inclusion of higher order terms in Margules' original power series [60].

*Importance of robust laboratory tests* The failure of the classical model to capture the laboratory observations presented in this work is only evident in either the retention profiles or when multiple impedance curves are treated simultaneously. A routine test measuring pressure drop and breakthrough concentration, particularly for short periods, may not be able to differentiate between filtration behaviour corresponding to a non-linear suspension function and classical filtration behaviour.

*General multicomponent particle models* The comparison in this study between the Margules and multicomponent particle filtration models is limited to the binary model. However, it has been shown that the binary model is very efficient at approximating the behaviour of multicomponent models with more particles [61]. This analysis extends both to the suspension function but also to typical measured data such as breakthrough concentration and retention profiles. Thus we can say that the Margules model approximates any *n*-particle population of injected particles.

Comparison between the binary and Margules models resulted in an explicit polynomial equation (32) that allows for determination of the Margules exponent, $A$, based on the parameters of the binary system. There are some issues with the polynomial approximation, particularly in the regions of $\theta \sim 0$, $c_1^0 \sim 0$, and $c_1^0 \sim 1$, where the binary model exhibits approximately linear behaviour. Near these boundaries, Khazali, Malgaresi, Russell, Osipov, Kuzmina and Bedrikovetsky [40] derived explicit expressions for the suspension function for a binary mixture. Using this explicit expression will allow for a much simpler process of comparing the binary and Margules models and as such the polynomial approximation may no longer be necessary.

## 7. Conclusions

Development of a modified deep bed filtration model based on a non-linear suspension function and comparison with multicomponent filtration models and experimental data allows drawing the following conclusions:

- Based on an analogy with chemical reactions, it is possible to derive a non-linear suspension function using the assumption of symmetric Gibbs free energy as used by Margules
- In the limit of the Margules exponent tending to zero, $A \rightarrow 0$, the system reverts to the classical model for colloidal filtration
- Travelling wave solutions of the Margules model and a simplified version with a linear filtration function shows good agreement when the Margules exponent is less than 0.4
- The simplified Margules model for deep bed filtration allows for an exact solution
- The Margules model approximates well the model for filtration of a binary mixture of colloids (which itself has been shown to approximate well any *n*-particle system)
- The Margules model contains 3 unknown parameters ($A, \lambda, s_m$) allowing for more stable optimization of laboratory data compared to the existing multicomponent model which has at minimum 4 unknown parameters
- An empirical relationship is derived for the optimal Margules exponent ($A$) which exhibits the closest agreement with the binary model for all possible parameter values ($\theta, c_{10}/c_{20}$)
- The new Margules model exhibits hyper-exponential retention profiles and shows good agreement with laboratory coreflooding data

Appendix A. Derivation of the activity coefficients of the Margules model

The Margules equation describes the excess Gibbs free energy of a solution of multiple components. The excess Gibbs free energy, $G^E$ is the difference between the actual Gibbs free energy, $G$, and that for an ideal solution, $G^{ideal}$

$$G = G^{ideal} + G^E \tag{A1}$$

For a mixture of $n$ components, the ideal Gibbs free energy is given by:

$$G^{ideal} = RT\sum_{i=1}^{n} x_i \ln(x_i) \tag{A2}$$

The total Gibbs free energy of a real mixture (non-ideal solution) can be written as:

$$G = RT\sum_{i=1}^{n} x_i \ln(x_i) + RT\sum_{i=1}^{n} x_i \ln(\gamma_i) \tag{A3}$$

Thus we find the following expression for the excess Gibbs free energy:

$$\frac{G^E}{RT} = \sum_{i=1}^{n} x_i \ln\gamma_i \tag{A4}$$

The two suffix Margules model relates the excess Gibbs free energy to the molar concentrations for each of the two components as follows [62]:

$$\frac{G^E}{RT} = Ax_1x_2 \tag{A5}$$

From equations (A4) and (A5), we conclude that:

$$\sum_{i=1}^{n} x_i \ln\gamma_i = Ax_1x_2 \tag{A6}$$

Also, we know that:

$$x_i = \frac{n_i}{n} \tag{A7}$$

In a binary system:

$$n = n_1 + n_2 \tag{A8}$$

Therefore, equation (A6) in a binary system becomes:

$$x_1 \ln\gamma_1 + x_2 \ln\gamma_2 = Ax_1x_2 \tag{A9}$$

Also, according to (A7):

$$n_1 \ln \gamma_1 + n_2 \ln \gamma_2 = Anx_1x_2 \tag{A10}$$

From (A10) we can conclude that:

$$\ln(\gamma_1) = \frac{\partial(Anx_1x_2)}{\partial n_1}|_{T,p,n_2} \tag{A11}$$

$$\ln(\gamma_1) = \frac{\partial\left(\dfrac{An_1n_2}{n_1+n_2}\right)}{\partial\left(n_1\right)} = An_2 \frac{n_1+n_2-n_1}{\left(n_1+n_2\right)^2} \tag{A12}$$

$$\ln(\gamma_1) = An_2\left(\frac{n_2}{(n_1+n_2)^2}\right) = An_2\left(\frac{n_2}{n^2}\right) = Ax_2^2 \tag{A13}$$

Therefore according to (A13) for the activity coefficients we will have:

$$\begin{aligned}\gamma_1 &= \exp(Ax_2^2)\\ \gamma_2 &= \exp(Ax_1^2)\end{aligned} \tag{A14}$$

Appendix B. Travelling wave solution using Margules suspension function

Here we consider the conditions of neglecting the exponential term of retained concentration *s*, i.e., for which values of *A* and injected concentration $c^0$ the form of filtration function in Eq. (16) is valid. Consider the system with the general Margules filtration rate derived in Section 2.2:

$$\frac{\partial\left(c+s\right)}{\partial t} + \frac{\partial c}{\partial x} = 0,\ \frac{\partial s}{\partial t} = \left(1-\frac{s}{s_m}\right)\exp\left[A\left(1-\frac{s}{s_m}\right)^2\right]c\exp\left[Ac^2\right]$$

(B1)

with initial and boundary conditions (10). Let us consider travelling wave (TW) solution of system (B1):

$$c\left(x,t\right)=C\left(y\right),\ s\left(x,t\right)=S\left(y\right),\ y=x-Dt \tag{B2}$$

Substitution of (B2) into Eqs. (B1) yields the following system of ODEs:

$$-D\frac{d(C+S)}{dy}+\frac{dC}{dy}=0,$$
$$-D\frac{dS}{dy}=\left(1-\frac{S}{s_m}\right)\exp\left[A\left(1-\frac{S}{s_m}\right)^2\right]C\exp\left[AC^2\right] \tag{B3}$$

Substitution of (B2) into initial and boundary conditions (10) yields

$$y\to\infty:\ C=S=0,$$
$$y\to-\infty:\ C=1,\ S=s_m \tag{B4}$$

Integrating the first ODE in Eq. (B3) in $y$ results in:

$$-D(C+S)+C=const \tag{B5}$$

Evaluating Eq. (B5) at $y$ near positive and negative infinity and accounting for boundary conditions (B4) allows calculating the constant and wave speed in Eq. (B5):

$$-D(C+S)+C=0=-D(1+s_m)+1,$$
$$D=\frac{1}{1+s_m}<1 \tag{B6}$$

Expressing $C$ vs $S$ from Eq. (B6):

$$C=\frac{S}{s_m} \tag{B7}$$

and substituting it into Eq. (B3) yields:

$$-Ds_m\frac{dC}{dy}=(1-C)\exp\left[A(1-C)^2\right]C\exp\left[AC^2\right] \tag{B8}$$

Equation (B8) is solved by separation of variables, resulting in an implicit expression for $C(y)$:

$$\int_1^C\frac{dz}{(1-z)\exp\left[A(1-z)^2\right]z\exp\left[Az^2\right]}=-\frac{s_m}{1+s_m}y \tag{B9}$$

Fig. 10 shows the solid integral trajectories given by Eq. (B9) for different values of constant $A$. For all trajectories, $C(y)$ starts from $C$=1 at minus infinity and tends to $C$=0 as $y$ tends to infinity. The integral trajectories expressed by the simplified Margules equation:

$$\int_1^C\frac{dz}{(1-z)z\exp\left[Az^2\right]}=-\frac{s_m}{1+s_m}y \tag{B10}$$

are shown by dotted curves.

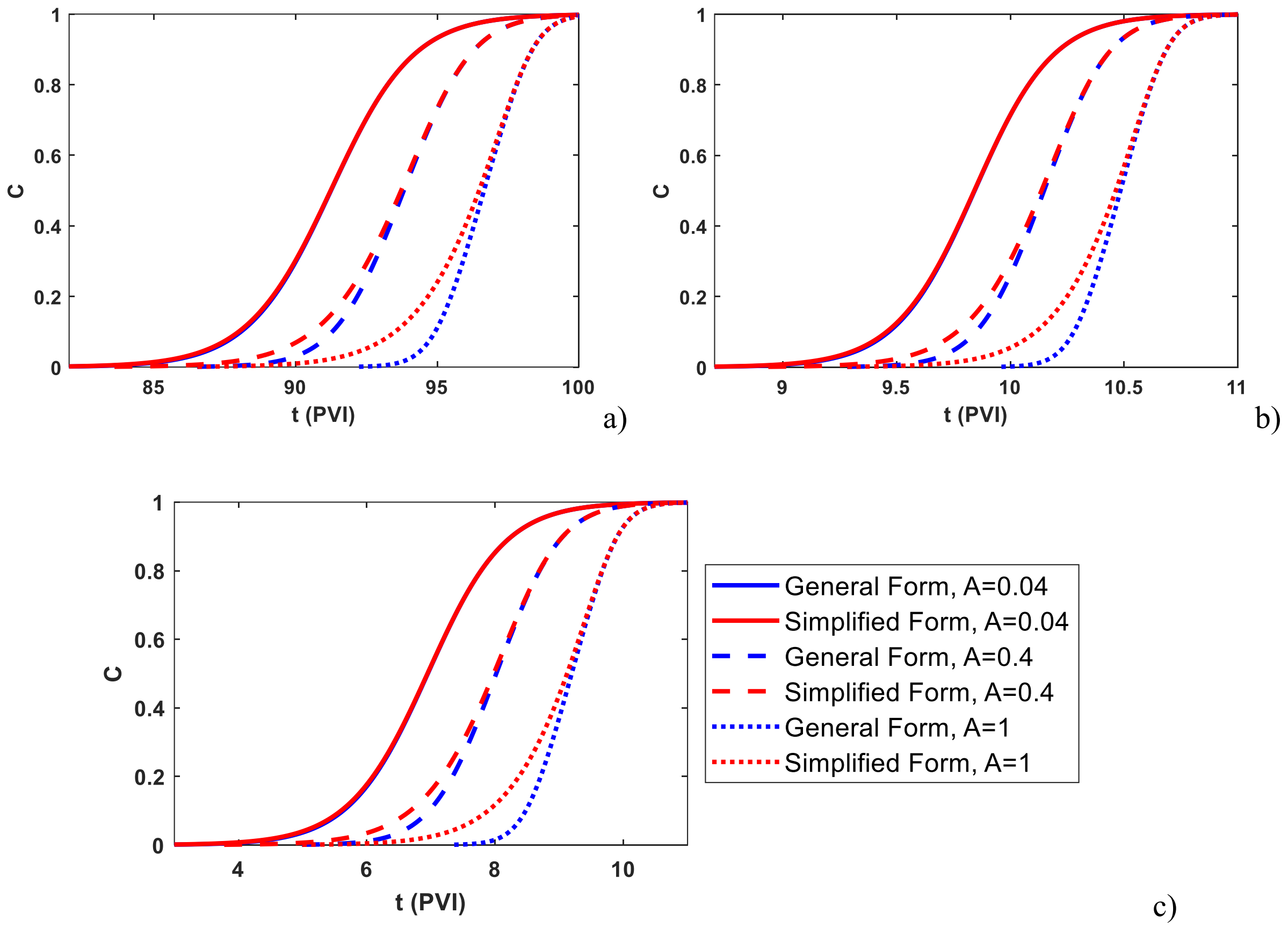

Fig. 10. Sensitivity analysis to compare the general and simplified form of Margules when a) $s_m$=100 and $\lambda$=70, b) $s_m$=10, and $\lambda$=70, c) $s_m$=10 and $\lambda$=20.

Consider the condition of neglecting the exponential term for $s$ in Eq. (14):

$$A\left(1-\frac{S}{s_m}\right)^2 << 1 \qquad \text{(B11)}$$

Accounting for relationship (B7) transforms the condition (B11) into

$$C >> 1-\frac{1}{\sqrt{A}} \qquad \text{(B12)}$$

corresponding to

$$y << y_A,\ \ y_A = -\frac{1+s_m}{s_m}\int_{1}^{1-\frac{1}{\sqrt{A}}} \frac{dz}{(1-z)\exp\left[A(1-z)^2\right] z\exp\left[Az^2\right]} \qquad \text{(B13)}$$

The values of $y_A$ are shown in Fig. 10 for corresponding integral curves.

In the $(x,t)$ plane, the condition (B13) corresponds to

$$x - Dt << y_A \tag{B14}$$

So while there is agreement at $A$=0.04 for all $t$, we would say that for $A$=0.4, there is agreement between the general and simplified Margules capture rate equations for $t$~7 PVI.

Appendix C. Upscaling of multi population colloidal transport

Substituting Eq. (18) into Eq. (17) yields:

$$\frac{\partial c_k}{\partial t} + \frac{\partial c_k}{\partial x} = -Bh\left(\sum_{i=1}^{m} s_i\right) F_k\left(c_k\right) \tag{C1}$$

Characteristic form of first order partial differential equations (PDE) is a system of ordinary differential equations along characteristic lines:

$$\frac{dt}{dx} = 1, \quad \frac{dc_k}{dx} = -Bh\left(\sum_{i=1}^{m} s_i\right) F_k\left(c_k\right) \tag{C2}$$

Rewriting equation (C2) in another form:

$$\frac{dG_k\left(c_k\right)}{dx} = -Bh\left(\sum_{i=1}^{m} s_i\right); \quad G_k\left(c_k\right) = \int_{c_k^o}^{c_k} \frac{du}{F_k(u)}, \; k = 1,2,...,N \tag{C3}$$

The right hand sides of equations (C3) coincide for all $k$=1,2…$N$.

Functions $G_k$ are equal at $x$=0. Therefore, values $G_k(c_k(x,t))$ are equal for all $k$. In particular,

$$G_k\left(c_k\right) = G_1(c_1) \tag{C4}$$

The total concentrations for suspended and retained particles and occupied area are determined as per Eq. (20).

Expressing individual concentrations $c_k$ from equation (C4) accounting for monotonicity of functions $G_k$:

$$c_k = G_k^{-1}\left[G_1(c_1)\right] \tag{C5}$$

Substituting relationships (C5) into equations (20) and expressing total concentration $c$ via $c_1$ yields:

$$c=\sum_{k=1}^{N}G_k^{-1}\left[G_1(c_1)\right],\ c_1=g_1(c),\ g_1=\left[\sum_{k=1}^{N}G_k^{-1}(G_1)\right]^{-1} \tag{C6}$$

Substituting Eqs. (C6) into Eq. (C5) leads to expressing each individual component $c_1$ versus total suspension concentration

$$c_k=G_k^{-1}\left[G_1(g_1(c))\right]=g_k(c) \tag{C7}$$

Substituting Eqs. (20) into Eq. (17):

$$\frac{\partial}{\partial t}\left(c+s\right)+\frac{\partial c}{\partial x}=0 \tag{C8}$$

Similarly, averaging over Eqs. (18) for $k$=1,2…$N$ yields the kinetics equation for retention rate.

$$\frac{\partial s}{\partial t}=f(c),\ f(c)=\sum_{k=1}^{N}F_k(g_k(c)) \tag{C9}$$

Expressing individual concentrations via total concentration in equations (C7) facilitates downscaling:

$$c_k(x,t)=c_k^0\left[\frac{g_1(c(x,t))}{c_1^0}\right]^{\frac{\lambda_k}{\lambda_1}},\ \ s_k(x,t)=\int_0^t F_k(c_k(x,y))dy \tag{C10}$$

Appendix D. Analytical solution for 1D colloidal transport with a Langmuir filtration function

Here we solve Eqs. (7) and (8) with Langmuir filtration function (5) and arbitrary suspension function. Initial and boundary conditions are given by Eq. (10).

Introduce a new variable:

$$\tau=t-x \tag{D1}$$

The unknowns in Eqs. (7) and (8) are $c(x,\tau(x,t))$ and $s(x,\tau(x,t))$. Recalculating the derivatives in Eq. (7) using the new coordinate system, $(x,\tau)$ results in:

$$\frac{\partial c}{\partial x}=-\frac{\partial s}{\partial \tau} \tag{D2}$$

Similarly, substituting Langmuir filtration function for the capture rate, given by Eq. (5), into Eq. (8) results in:

$$\frac{\partial s}{\partial \tau}=\left(1-\frac{s}{s_m}\right)f\left(c\right) \tag{D3}$$

Substituting RHS of Eq. (D3) into Eq. (D2), we obtain:

$$\frac{\partial c}{\partial x} = -\left(1 - \frac{s}{s_m}\right) f(c) \tag{D4}$$

The dimensionless initial conditions for the new system ((D2),(D3)) are:

$$\tau = -x: \ c = s = 0 \tag{D5}$$

The dimensionless boundary condition at $x=0$ is:

$$x = 0: \ c = 1 \tag{D6}$$

Ahead of the injected particle front we have:

$$c = s = 0 \tag{D7}$$

For the first injection period, the suspension concentration behind the front $c^-$ can be obtained substituting $s=0$ into Eq. (D4):

$$\frac{\partial c}{\partial x} = -f(c) \tag{D8}$$

Introduce a new function:

$$\varphi(1) - \varphi\left(c^-(x,0)\right) = x, \quad \varphi'(c) = \frac{1}{f(c)} \tag{D9}$$

Eq. (D4) can be rewritten in the form:

$$\frac{1}{f(c)} \frac{\partial c}{\partial x} + 1 = \frac{s}{s_m} \tag{D10}$$

Taking $x$-derivative from both sides of Eq. (D10) yields:

$$\frac{\partial \varphi(c)}{\partial x} = \frac{1}{f(c)} \frac{\partial c}{\partial x} = \frac{\partial \varphi(c)}{\partial c} \frac{\partial c}{\partial x} \tag{D11}$$

Expressing $s(x,\tau)$ from Eq. (D10) results in:

$$s(x,\tau) = s_m \left[ \frac{1}{f(c)} \frac{\partial c}{\partial x} + 1 \right] = s_m \left[ \frac{\partial \varphi(c)}{\partial x} + 1 \right] \tag{D12}$$

Substituting Eq. (D12) into Eq. (D2):

$$\frac{\partial}{\partial \tau}\left(s_m \frac{\partial \varphi(c)}{\partial x}\right)+\frac{\partial c}{\partial x}=0 \tag{D13}$$

Changing the order of derivatives in first term of LHS of Eq. (D13), integrating in $x$, accounting for boundary condition (D6), yields:

$$s_m \frac{\partial \varphi(c)}{\partial \tau}+c=1 \tag{D14}$$

For any fixed $x>0$, Eq. (D14) is solved by separation of variables:

$$\frac{\partial c}{(1-c) f(c)}=\frac{\partial \tau}{s_m} \tag{D15}$$

Integrating both sides of Eq. (D15) from $c^{-}(x)$ to $c(x,\tau)$ yields:

$$\int_{c^{-}(x)}^{c} \frac{du}{(1-u) f(u)}=\frac{\tau}{s_m} \tag{D16}$$

Taking derivative in $x$ of both sides of Eq. (D16) leads to:

$$\frac{1}{(1-c) f(c)} \frac{\partial c}{\partial x}-\frac{1}{\left(1-c^{-}(x,0)\right) f\left(c^{-}(x,0)\right)} \frac{dc^{-}(x,0)}{dx}=0 \tag{D17}$$

From substituting Eq. (D12) into Eq. (D17) we can determine $s(x,\tau)$.

$$\frac{s(x,\tau)-s_m}{1-c(x,\tau)}=\frac{-s_m}{(1-c(x,0))} \tag{D18}$$

Here in the first region, the initial condition is: $s(x,0)=0$. Rearranging and returning to $(x,t)$ coordinates yields Eq. (26).

# Supplementary Material

## Section S1. Matching the binary and Margules models using BTC and RPs

In this section we present the results of matching the binary and Margules models for non-linear suspension functions. Here the matching is performed by using each model to calculate the breakthrough concentration and retention profile after some fixed time. The Margules model is then tuned such that the outputs using each suspension function is minimised (according to Eq. (33)). The results are performed over a range of values for the binary model ($\theta$, $c_1^0$). The binary model exhibits non-linear behaviour when the difference between filtration coefficients is high (high $\theta$) and there are abundant amounts of each type of particle ($c_1^0$ not near 0 or 1). This coincides with higher values of $A$ as shown in Table S1. The optimal values of the filtration coefficient show higher values when $\theta$ is high and a high proportion of particles have a higher filtration coefficient ($c_1^0$ near 1). The values of $R^2$ for the two datasets show good agreement.

Table S1. Tuned values of $A$ obtained by matching the binary and Margules breakthrough concentration and retention profiles (Langmuir filtration function)

| $\theta$ \ $c_1^0$ | 0 | 0.1 | 0.2 | 0.3 | 0.4 | 0.5 | 0.6 | 0.7 | 0.8 | 0.9 | 1 |
|---|---|---|---|---|---|---|---|---|---|---|---|
| 1 | 0 | 2.2665 | 2.8866 | 2.9593 | 3.1822 | 3.2142 | 3.3143 | 3.4011 | 7.8248 | 2.8869 | 0.0000 |
| 0.9 | 0 | 0.9305 | 1.3464 | 1.4895 | 1.5232 | 1.5113 | 1.4475 | 1.3779 | 1.2013 | 0.9421 | 0.0000 |
| 0.8 | 0 | 0.4326 | 0.6605 | 0.7369 | 0.7504 | 0.7205 | 0.6621 | 0.9947 | 0.4600 | 0.2971 | 0.0000 |
| 0.7 | 0 | 0.2287 | 0.3491 | 0.4072 | 0.4199 | 0.4034 | 0.3655 | 0.3099 | 0.2378 | 0.1493 | 0.0000 |
| 0.6 | 0 | 0.1321 | 0.1990 | 0.2377 | 0.2517 | 0.2467 | 0.2248 | 0.1954 | 0.1549 | 0.1047 | 0.0000 |
| 0.5 | 0 | 0.0821 | 0.1200 | 0.1433 | 0.1545 | 0.1554 | 0.1477 | 0.1326 | 0.1106 | 0.0834 | 0.0000 |
| 0.4 | 0 | 0.0549 | 0.0725 | 0.0888 | 0.0936 | 0.0988 | 0.0936 | 0.0904 | 0.0771 | 0.0682 | 0.0000 |
| 0.3 | 0 | 0.0400 | 0.0501 | 0.0572 | 0.0618 | 0.0639 | 0.0639 | 0.0619 | 0.0604 | 0.0544 | 0.0000 |
| 0.2 | 0 | 0.0319 | 0.0335 | 0.0394 | 0.0383 | 0.0431 | 0.0436 | 0.0434 | 0.0389 | 0.0408 | 0.0000 |
| 0.1 | 0 | 0.0280 | 0.0292 | 0.0302 | 0.0310 | 0.0316 | 0.0321 | 0.0323 | 0.0325 | 0.0325 | 0.0000 |
| 0 | 0 | 0 | 0 | 0 | 0 | 0 | 0 | 0 | 0 | 0 | 0 |

Table S2. Tuned values of $\lambda$ obtained by matching the binary and Margules breakthrough concentration and retention profiles (Langmuir filtration function)

| $\theta$ \ $c_1^0$ | 0 | 0.1 | 0.2 | 0.3 | 0.4 | 0.5 | 0.6 | 0.7 | 0.8 | 0.9 | 1 |
|---|---|---|---|---|---|---|---|---|---|---|---|
| 1 | 0.6 | 0.1927 | 0.2094 | 0.3433 | 0.4042 | 0.5475 | 0.7031 | 0.9728 | 0.9966 | 2.1813 | 20 |
| 0.9 | 0.6 | 0.4087 | 0.4146 | 0.5018 | 0.6338 | 0.8063 | 1.0384 | 1.3462 | 1.8065 | 2.6104 | 8.478 |
| 0.8 | 0.6 | 0.5352 | 0.5582 | 0.6416 | 0.7637 | 0.9237 | 1.1252 | 1.3814 | 1.7173 | 2.1955 | 3.0097 |
| 0.7 | 0.6 | 0.5859 | 0.6217 | 0.6858 | 0.7755 | 0.8900 | 1.0301 | 1.1987 | 1.4020 | 1.6504 | 1.9656 |
| 0.6 | 0.6 | 0.6035 | 0.6403 | 0.6898 | 0.7547 | 0.8308 | 0.9239 | 1.0273 | 1.1489 | 1.2887 | 1.4501 |
| 0.5 | 0.6 | 0.6073 | 0.6388 | 0.6775 | 0.7231 | 0.7755 | 0.8348 | 0.9010 | 0.9747 | 1.0564 | 1.1468 |
| 0.4 | 0.6 | 0.6052 | 0.6328 | 0.6593 | 0.6939 | 0.7268 | 0.7682 | 0.8075 | 0.8573 | 0.9011 | 0.9800 |
| 0.3 | 0.6 | 0.6005 | 0.6192 | 0.6393 | 0.6609 | 0.6839 | 0.7084 | 0.7344 | 0.7604 | 0.7895 | 0.8571 |
| 0.2 | 0.6 | 0.5949 | 0.6092 | 0.6196 | 0.6354 | 0.6463 | 0.6604 | 0.6750 | 0.6932 | 0.7058 | 0.7500 |
| 0.1 | 0.6 | 0.5891 | 0.5949 | 0.6008 | 0.6068 | 0.6130 | 0.6192 | 0.6255 | 0.6319 | 0.6384 | 0.6600 |
| 0 | 0.6 | 0.6 | 0.6 | 0.6 | 0.6 | 0.6 | 0.6 | 0.6 | 0.6 | 0.6 | 0.6 |

Table S3. Obtained values of $R_c^2$ obtained by matching the binary and Margules breakthrough concentration and retention profiles (Langmuir filtration function)

| $\theta$ \ $c_1^0$ | 0 | 0.1 | 0.2 | 0.3 | 0.4 | 0.5 | 0.6 | 0.7 | 0.8 | 0.9 | 1 |
|---|---|---|---|---|---|---|---|---|---|---|---|
| 1 | 1.0000 | 0.9488 | 0.9689 | 0.9852 | 0.9997 | 0.9985 | 0.9960 | 0.9967 | 0.9897 | 0.9614 | 1.0000 |
| 0.9 | 1.0000 | 0.9879 | 0.9839 | 0.9915 | 0.9964 | 0.9980 | 0.9975 | 0.9947 | 0.9871 | 0.9601 | 1.0000 |
| 0.8 | 1.0000 | 0.9961 | 0.9917 | 0.9911 | 0.9933 | 0.9958 | 0.9973 | 0.9979 | 0.9974 | 0.9965 | 1.0000 |
| 0.7 | 1.0000 | 0.9981 | 0.9961 | 0.9942 | 0.9929 | 0.9929 | 0.9937 | 0.9947 | 0.9955 | 0.9959 | 1.0000 |
| 0.6 | 1.0000 | 0.9988 | 0.9979 | 0.9966 | 0.9958 | 0.9942 | 0.9939 | 0.9925 | 0.9923 | 0.9921 | 1.0000 |
| 0.5 | 1.0000 | 0.9991 | 0.9985 | 0.9979 | 0.9971 | 0.9963 | 0.9955 | 0.9945 | 0.9934 | 0.9924 | 1.0000 |
| 0.4 | 1.0000 | 0.9992 | 0.9992 | 0.9986 | 0.9986 | 0.9977 | 0.9977 | 0.9967 | 0.9972 | 0.9953 | 1.0000 |
| 0.3 | 1.0000 | 0.9994 | 0.9992 | 0.9990 | 0.9988 | 0.9986 | 0.9983 | 0.9980 | 0.9976 | 0.9973 | 1.0000 |
| 0.2 | 1.0000 | 0.9994 | 0.9995 | 0.9992 | 0.9994 | 0.9990 | 0.9989 | 0.9988 | 0.9990 | 0.9985 | 1.0000 |
| 0.1 | 1.0000 | 0.9995 | 0.9994 | 0.9994 | 0.9994 | 0.9993 | 0.9993 | 0.9992 | 0.9992 | 0.9991 | 1.0000 |
| 0 | 1.0000 | 0.9994 | 0.9994 | 0.9994 | 0.9994 | 0.9994 | 0.9994 | 0.9994 | 0.9994 | 0.9994 | 1.0000 |

Table S4. Obtained values of $R_s^2$ obtained by matching the binary and Margules breakthrough concentration and retention profiles (Langmuir filtration function)

| $\theta$ \ $c_1^0$ | 0 | 0.1 | 0.2 | 0.3 | 0.4 | 0.5 | 0.6 | 0.7 | 0.8 | 0.9 | 1 |
|---|---|---|---|---|---|---|---|---|---|---|---|
| 1 | 1.0000 | 0.7428 | 0.7188 | 0.7072 | 0.6898 | 0.6934 | 0.6750 | 0.6978 | 0.8409 | 0.6878 | 1.0000 |
| 0.9 | 1.0000 | 0.9900 | 0.9856 | 0.9747 | 0.9638 | 0.9544 | 0.9469 | 0.9430 | 0.9417 | 0.9464 | 1.0000 |
| 0.8 | 1.0000 | 0.9989 | 0.9974 | 0.9933 | 0.9882 | 0.9837 | 0.9807 | 0.9792 | 0.9790 | 0.9788 | 1.0000 |
| 0.7 | 1.0000 | 0.9996 | 0.9992 | 0.9981 | 0.9965 | 0.9946 | 0.9928 | 0.9913 | 0.9899 | 0.9882 | 1.0000 |
| 0.6 | 1.0000 | 0.9996 | 0.9996 | 0.9994 | 0.9988 | 0.9984 | 0.9975 | 0.9967 | 0.9956 | 0.9941 | 1.0000 |
| 0.5 | 1.0000 | 0.9996 | 0.9997 | 0.9997 | 0.9996 | 0.9994 | 0.9992 | 0.9989 | 0.9984 | 0.9977 | 1.0000 |
| 0.4 | 1.0000 | 0.9996 | 0.9995 | 0.9996 | 0.9995 | 0.9996 | 0.9994 | 0.9994 | 0.9989 | 0.9991 | 1.0000 |
| 0.3 | 1.0000 | 0.9995 | 0.9996 | 0.9996 | 0.9996 | 0.9996 | 0.9995 | 0.9995 | 0.9995 | 0.9994 | 1.0000 |
| 0.2 | 1.0000 | 0.9995 | 0.9994 | 0.9995 | 0.9994 | 0.9995 | 0.9995 | 0.9995 | 0.9993 | 0.9995 | 1.0000 |
| 0.1 | 1.0000 | 0.9995 | 0.9995 | 0.9995 | 0.9995 | 0.9995 | 0.9995 | 0.9995 | 0.9995 | 0.9995 | 1.0000 |
| 0 | 1.0000 | 0.9995 | 0.9995 | 0.9995 | 0.9995 | 0.9995 | 0.9995 | 0.9995 | 0.9995 | 0.9995 | 1.0000 |

The data in Table S1 is tuned using polynomial (32) to allow for an explicit equation to transfer between the binary and Margules models. Contours of this plot are given in Fig. S1 and values of the coefficients are presented in Table S5.

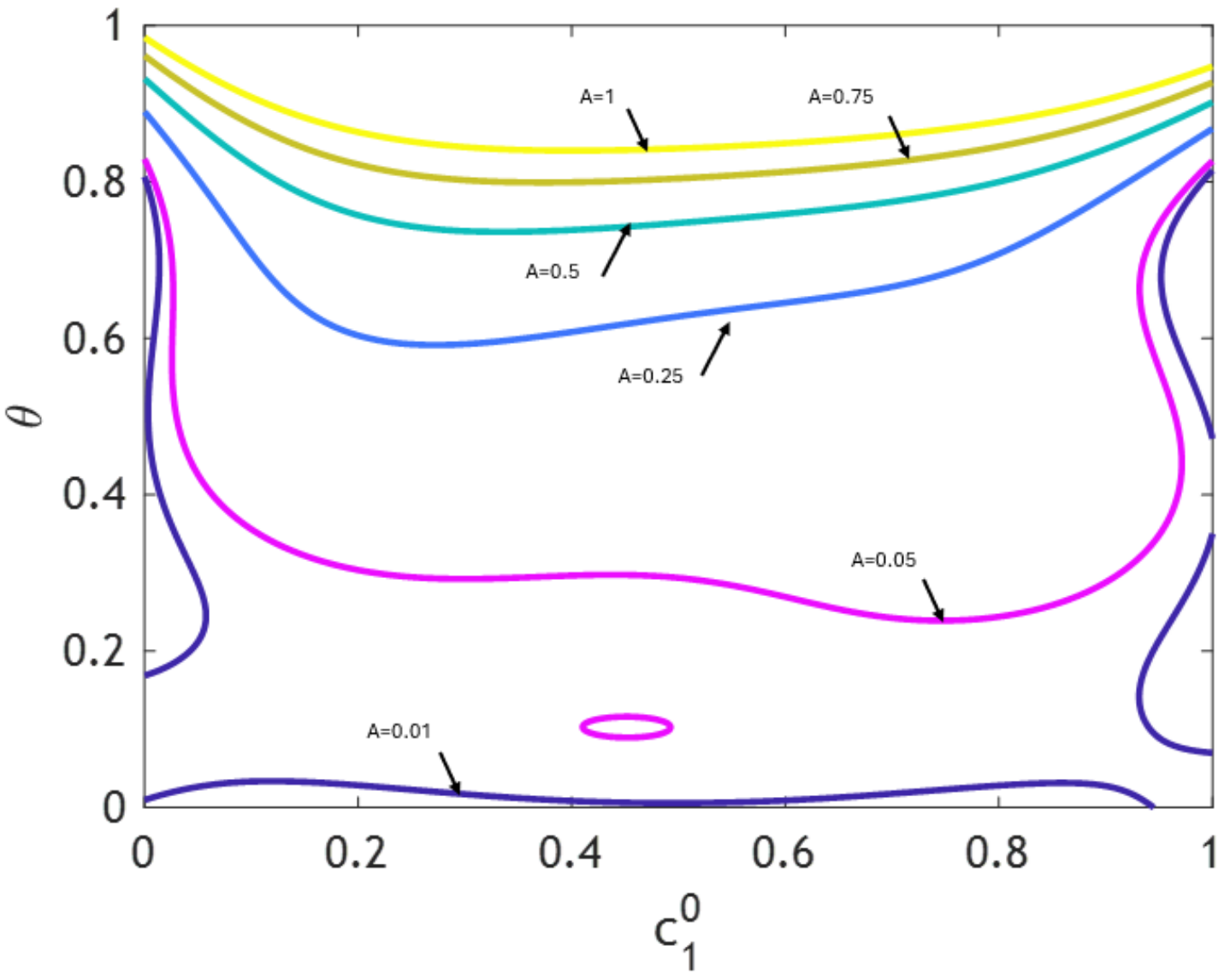

Fig. S1. Contours of polynomial **Error! Reference source not found.** for $A$ obtained by matching the binary and Margules breakthrough concentration and retention profiles (Langmuir filtration function)

Table S5. Coefficients of the polynomial fit for $A$ obtained by matching the binary and Margules breakthrough concentration and retention profiles (Langmuir filtration function)

| Coefficient | Values | Coefficient | Values |
|---|---|---|---|
| $P_{00}$ | 0 | $P_{41}$ | -11.894 |
| $P_{10}$ | -0.4372 | $P_{14}$ | 10.239 |
| $P_{01}$ | 1.1322 | $P_{42}$ | 0 |
| $P_{11}$ | 3.4297 | $P_{24}$ | 0 |
| $P_{20}$ | 2.4586 | $P_{43}$ | 0 |
| $P_{02}$ | -11.721 | $P_{34}$ | 0 |
| $P_{21}$ | -14.654 | $P_{44}$ | 0 |
| $P_{12}$ | 0.1696 | $P_{50}$ | -0.7495 |
| $P_{22}$ | 3.3593 | $P_{05}$ | 29.385 |
| $P_{30}$ | -4.6611 | $P_{51}$ | 0 |
| $P_{03}$ | 40.811 | $P_{15}$ | 0 |
| $P_{31}$ | 21.877 | $P_{52}$ | 0 |
| $P_{13}$ | -2.1042 | $P_{25}$ | 0 |
| $P_{32}$ | 4.4098 | $P_{53}$ | 0 |
| $P_{23}$ | -14.19 | $P_{35}$ | 0 |
| $P_{33}$ | 0 | $P_{54}$ | 0 |
| $P_{40}$ | 3.4181 | $P_{45}$ | 0 |
| $P_{04}$ | -58.418 | $P_{55}$ | 0 |

## Section S2. Matching experimental data

Fig. S2 shows another three impedance curves taken from Ramachandran and Fogler [1] with the corresponding tuned parameters presented in Table S6. The curves in Fig. S2 show that the binary model exhibits the best fit, while the Margules model also follows the trend of the data well. The classical model cannot simultaneously fit the two curves used for tuning ($c^0$=1235 ppm and $c^0$=720 ppm) and consequently fails to predict the behaviour at $c^0$=220 ppm. The other two models perform this prediction well, indicating that they are capturing the underlying retention behaviour of the system.

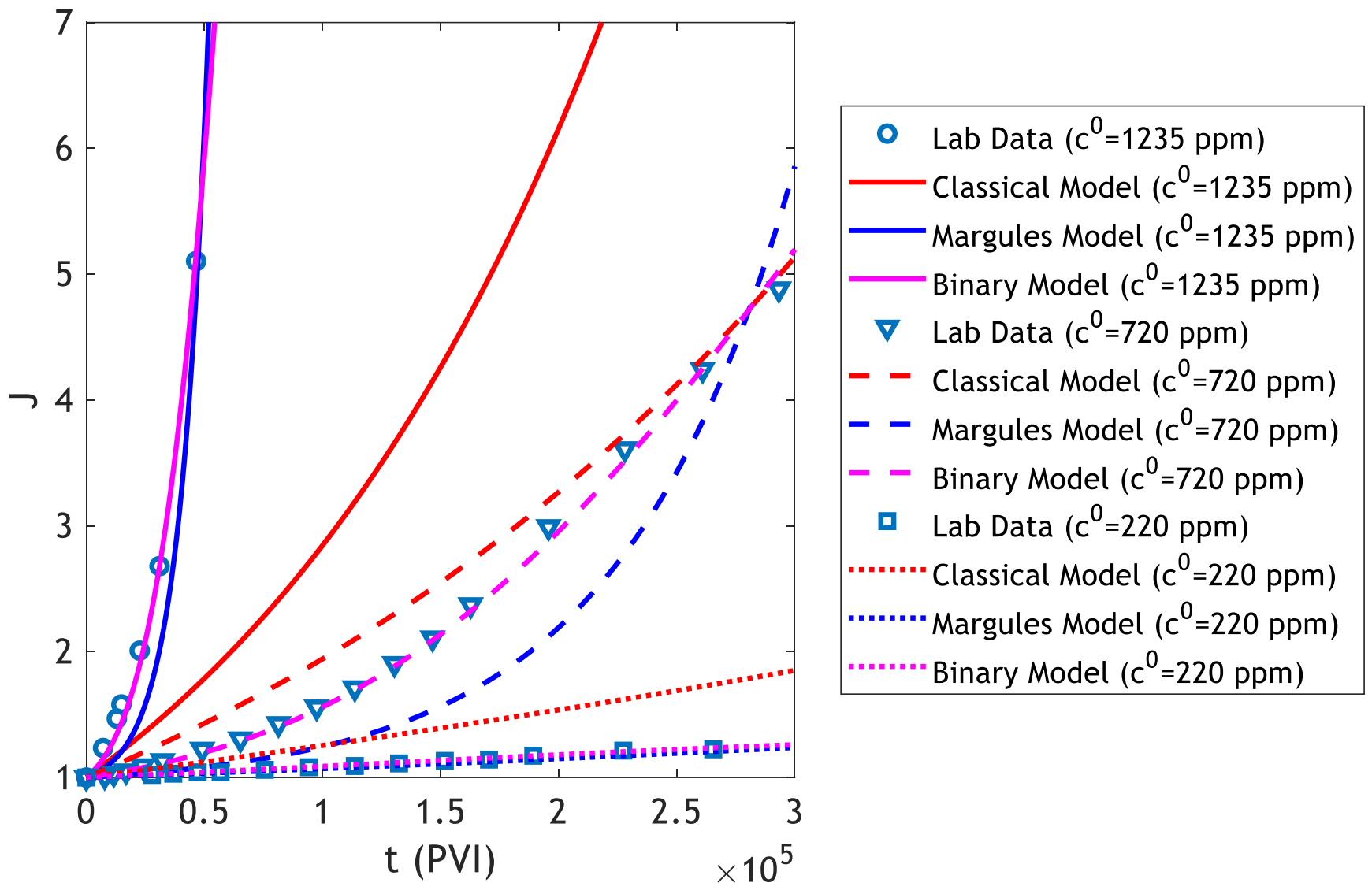

Fig. S2. Treatment of impedance lab data taken from Ramachandran and Fogler [1]

Table S6. Tuning parameters for treatment of impedance lab data taken from Ramachandran and Fogler [1] (Fig. S4)

| Models | Total injected concentration $c^0$ | $A\times10^{-5}$ | $\beta$ | $c_1^0$ [ppm] | $s_m$ [$L^{-3}$] | $\lambda_1$ [$m^{-1}$] | $\lambda_2$ [$m^{-1}$] | $R_p^2$ | Dataset |
|---|---|---|---|---|---|---|---|---|---|
| | $c^0$= 1235 ppm | | | | | | | - | Tuning |
| Classical | $c^0$= 720 ppm | | $5.02\times10^5$ | | $-4.61\times10^{-6}$ | $2.70\times10^{-1}$ | | 0.957 | Tuning |
| | $c^0$= 220 ppm | | | | | | | - | Prediction |
| | $c^0$= 1235 ppm | 14.339 | | | | | | 0.911 | Tuning |
| Margules | $c^0$= 720 ppm | 17.747 | $2.77\times10^6$ | | $-3.08\times10^{-8}$ | $2.85\times10^{-3}$ | | 0.861 | Tuning |
| | $c^0$= 220 ppm | 227.273 | | | | | | 0.899 | Prediction |
| | $c^0$= 1235 ppm | | | 203.282 | | | | 0.995 | Tuning |
| Binary | $c^0$= 720 ppm | | $1.07\times10^6$ | 184.393 | $-1.52\times10^{-7}$ | $4.55\times10^{-2}$ | $5.50\times10^{-3}$ | 0.998 | Tuning |
| | $c^0$= 220 ppm | | | 52.7511 | | | | 0.992 | Prediction |

Fig. S3 shows the laboratory data on injection of spherical fluorescent carboxylate-modified polystyrene latex particles with diameter 1.0 μm into quartz sand packs with grain sizes 417-600 μm with length 20 cm, diameter 3.81 cm and salinity 1.0 mM for NaHCO3 [2]. The tuning parameters and $R^2$ values are presented in Table S7. The traditional model captures the breakthrough concentration well but cannot match the hyperexponential retention profiles (Fig. S3b) while the binary and Margules models can. The binary model shows a significantly better fit to the retention profile compared to the Margules model.

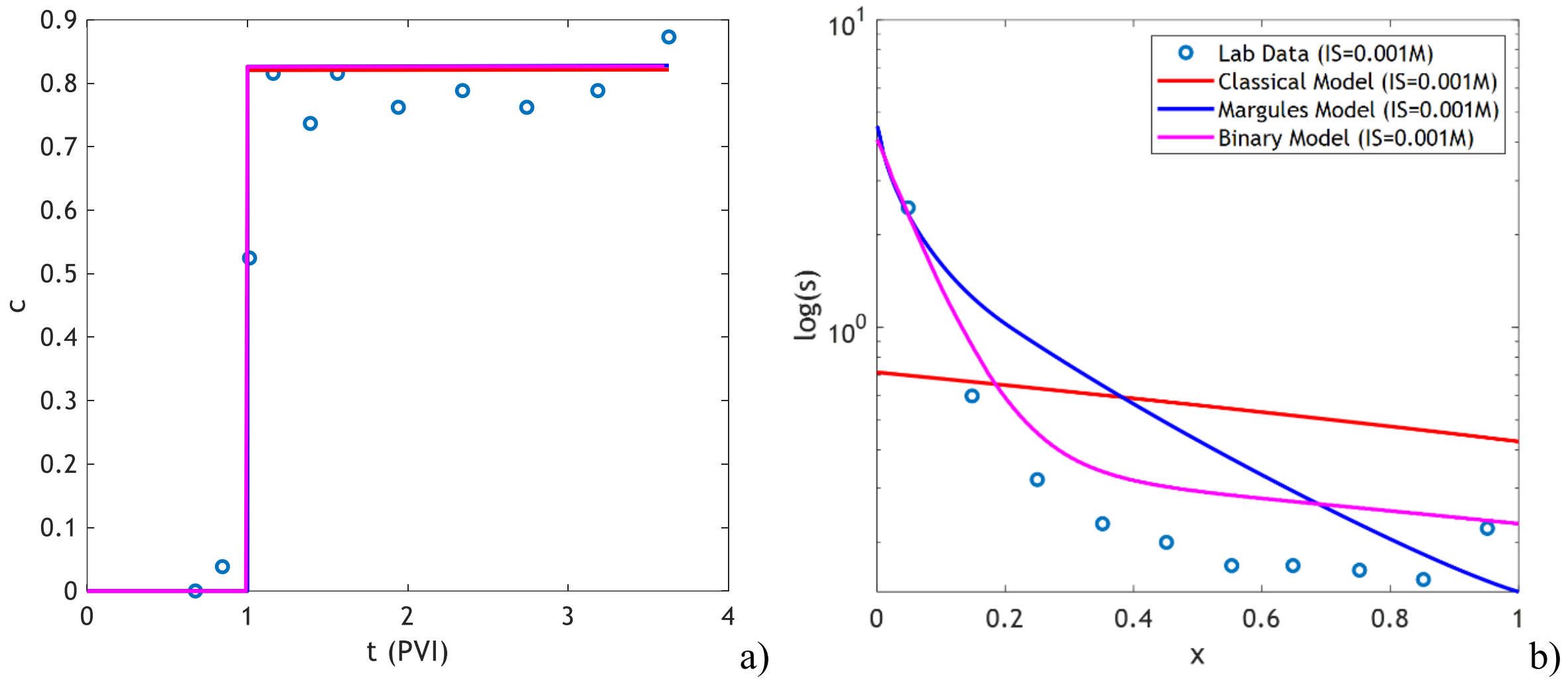

Fig. S3. Simultaneous treatment of the suspended and retained concentrations of lab data taken from Li and Johnson [2]

Table S7. Tuning parameters for treatment of breakthrough concentration and retention profiles data taken from Li and Johnson [2] (Fig. S2)

| Models | Test Conditions | $A \times 10^{-5}$ | $c_1^0$ [ppm] | $s_m$[$L^{-3}$] | $\lambda_1$ [$m^{-1}$] | $\lambda_2$ [$m^{-1}$] | $R_c^2$ | $R_s^2$ |
|---|---|---|---|---|---|---|---|---|
| Classical | $c^0$= 193 ppm<br>IS= 0.001 M | | | $6.40\times10^{-3}$ | 0.988 | | 0.89 | 0.125 |
| Margules | | $2.68\times10^{3}$ | | $9.81\times10^{-4}$ | $3.43\times10^{-4}$ | | 0.884 | 0.761 |
| Binary | | | 15.493 | $3.70\times10^{-3}$ | 67.599 | 0.536 | 0.884 | 0.962 |

Fig. S4 shows another series of test taken from the work by Yang, Bradford, Wang, Sharma, Shang and Li [3] with fitting parameters given in Table S8. In this case we see some increase in the breakthrough concentration with time, which is only captured by the binary and Margules models. The retention profiles are approximately exponential (linear on the *log(s) vs x* plot) and thus all three models exhibit good agreement.

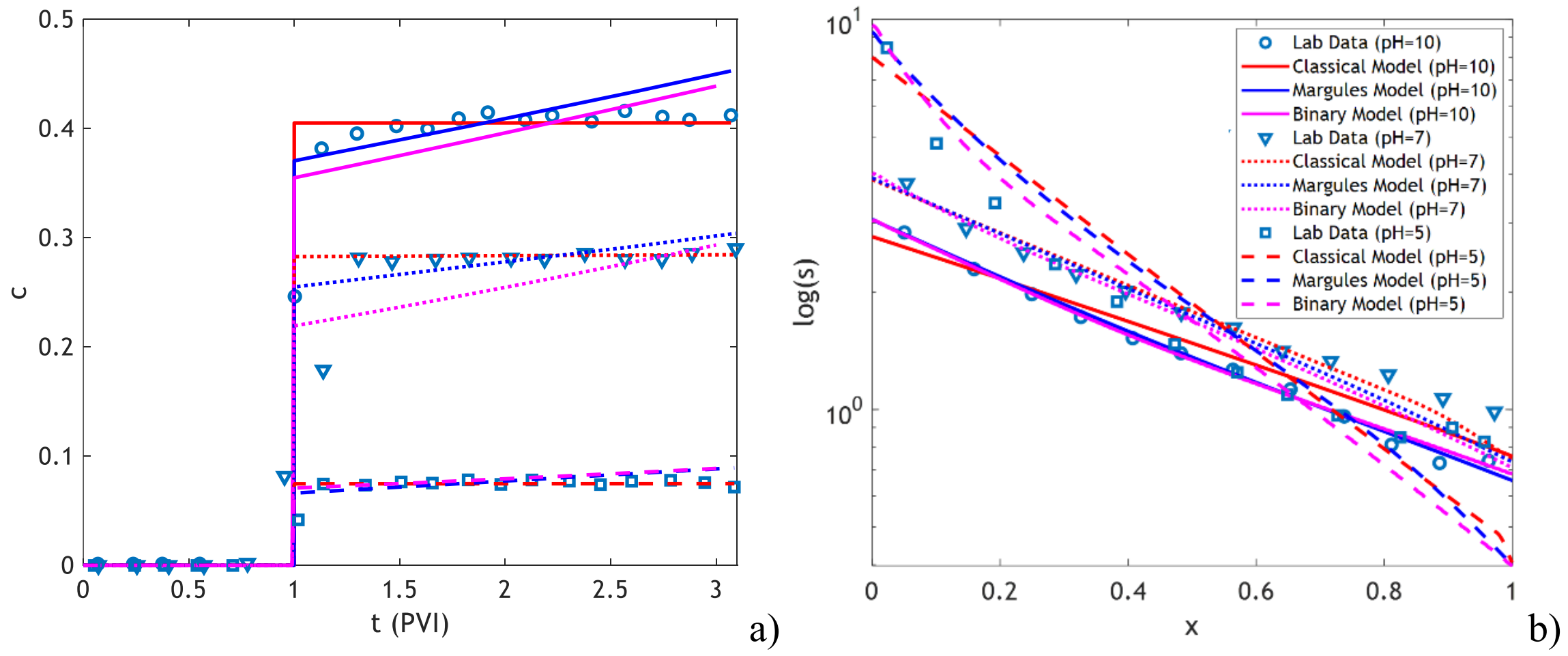

Fig. S4. Simultaneous treatment of the suspended and retained concentrations of lab data taken from Yang, Bradford, Wang, Sharma, Shang and Li [3]

Table S8. Tuning parameters for treatment of breakthrough concentration and retention profiles data taken from Yang, Bradford, Wang, Sharma, Shang and Li [3] (Fig. S3)

| Models | Test Conditions | $A\times10^{-5}$ | $c_1^0$ [ppm] | $s_m$[L$^{-3}$] | $\lambda_1$ [m$^{-1}$] | $\lambda_2$ [m$^{-1}$] | $R_c^2$ | $R_s^2$ |
|---|---|---|---|---|---|---|---|---|
| Classical | $c^0$= 50 ppm | | | $9.25\times10^{-2}$ | 7.529 | | 0.948 | 0.959 |
| Margules | IS = 0.001 M | 2980.4 | | $1.18\times10^{-4}$ | 6.078 | | 0.960 | 0.996 |
| Binary | BSA, pH=10 | | 5.6 | $1.23\times10^{-4}$ | 53.124 | 7.641 | 0.966 | 0.997 |
| Classical | $c^0$= 50 ppm | | | $7.60\times10^{-3}$ | 10.507 | | 0.938 | 0.974 |
| Margules | IS = 0.001 M | 791.6 | | $2.63\times10^{-4}$ | 10.660 | | 0.950 | 0.970 |
| Binary | BSA, pH=7 | | 2.7 | $1.76\times10^{-4}$ | 84.744 | 12.173 | 0.94 | 0.969 |
| Classical | $c^0$= 50 ppm | | | $9.84\times10^{-2}$ | 21.611 | | 0.945 | 0.892 |
| Margules | IS = 0.001 M | 2119.6 | | $4.22\times10^{-4}$ | 20.677 | | 0.935 | 0.911 |
| Binary | BSA, pH=5 | | 6.035 | $4.82\times10^{-4}$ | 148.029 | 20.983 | 0.924 | 0.958 |